\documentclass[twocolumn,english,prl,showpacs]{revtex4}
\usepackage[colorlinks=true,urlcolor=blue,citecolor=blue,linkcolor=blue]{hyperref}
\usepackage[sort&compress]{natbib}
\usepackage[T1]{fontenc}
\usepackage[latin9]{inputenc}
\usepackage{amssymb}
\usepackage{graphicx}
\usepackage{amsmath,color}
\usepackage{mathrsfs}
\usepackage{float}
\usepackage{indentfirst}
\usepackage{babel}
\usepackage{bbold}

\begin{document}

\title{Seeing Hofstadter's Butterfly in Atomic Fermi Gases}
%\title{Thermodynamic Signatures of Hofstadter's Butterfly in Optical Lattice}
\author{Lei Wang and Matthias Troyer}

\affiliation{Theoretische Physik, ETH Zurich, 8093 Zurich, Switzerland}

%\affiliation{$^{2}$Beijing National Lab for Condensed Matter Physics and Institute
%of Physics, Chinese Academy of Sciences, Beijing 100190, China}

\begin{abstract}
We propose a novel way to detect the fractal energy spectrum of the Hofstadter model from the density distributions of ultracold fermions in an external trap. At low temperature, the local compressibility is proportional to the density of states of the system which reveals the fractal energy spectrum. However, thermal broadening and noises in the real experimental situation inevitably smear out fine features in the density distribution. To overcome this difficulty, we use the maximum entropy method to extract the density of states directly from the noisy thermal density distributions. Simulations show that one is able to restore the core feature of the Hofstadter's butterfly spectrum with current experimental techniques. By further reducing the noise or the temperature, one can refine the resolution and observe fine structures of the butterfly spectrum. 
\end{abstract}

\pacs{67.85.Lm, 51.35.+a, 71.20.-b, 73.43.-f}

% 67.85.Lm degenerate Fermi gases
% 51.35.+a Compressibility
% 71.20.-b Electron density of states crystalline solids
% 73.43.-f Quantum Hall effects

\maketitle

%\begin{bibunit}[apsrev4-1]

%{\bf Introduction}
%Hofstadter model, fractal energy spectrum
The Hofstadter model~\cite{Hofstadter:1976wt} describes electrons moving in a 2D lattice exposed to a uniform magnetic field, where the interplay of lattice potential and magnetic field leads to an intriguing fractal energy spectrum, called Hofstadter's butterfly. Being one of the first quantum fractals discovered in nature, Hofstadter studied it around the same time when Mandelbrot coined the  term ``fractal''. 

%Solids state measurements 
Despite its mathematical beauty, Hofstadter's butterfly remained elusive for decades, because it requires infeasibly strong magnetic fields to see in the conventional crystals. Attempts have thus been made in  artificial superlattices, where much smaller magnetic fields  suffice. Early experiments reported  signatures for the fractal spectrum in the 2D electron gas with a weak lateral superlattice potential~\cite{Albrecht:2001kz} and recently more evidence was reported for  graphene superlattices~\cite{Dean:2013bv, Hunt:2013ef,Ponomarenko:2013hlb}. Since the system realizes a quantum Hall insulator when the chemical potential is in the energy gaps~\cite{Anonymous:4vo2mOrm}, these solid state experiments utilize the Hall conductance as a probe~\cite{Osadchy:2001jm} of the butterfly.

%Cold atom realization
Hofstadter's butterfly is also a long sought goal~\cite{Mueller:2004hc,Umucallar:2008fq,Gerbier:2010ho, Kolovsky:2011dv, Creffield:2013gp, Aidelsburger:2011hl, Aidelsburger:2013du} in cold atoms systems ever since the original proposal~\cite{Jaksch:2003gd}. Cold atomic gases offer a unique chance to study the model in the absence of disorder and with tunable interactions. Recently, two groups reported the realization of Hofstadter's model in optical lattices~\cite{Aidelsburger:2013ti, Miyake:2013vbc}, using laser-assisted tunneling to imprint complex phases to the hopping amplitudes and verifying the induced flux by studying the dynamics of bosons in the lattice. A natural next goal is the  definite observation of Hofstadter's butterfly in an optical lattice. However, contrary to solid state setups~\cite{Albrecht:2001kz,Dean:2013bv, Hunt:2013ef,Ponomarenko:2013hlb}, measuring the Hall conductance of the ultracold Fermi gases is not straightforward~\cite{Alba:2011cb,Goldman:2013dg,Anonymous:2013cs,Anonymous:2013jga,Liu:2013wy}. 

In this Letter, we thus propose a simple and novel way to measure Hofstadter's butterfly from the simplest thermodynamic quantity, the density distribution of the trapped Fermi gases. At low temperature the local compressibility is equal to the density of states (DOS), which directly reveals the fractal energy spectrum. %Previously, the global compressibility has been used to detect the Mott transition of fermi Hubbard model in optical lattices~\cite{Jordens:2008im, Schneider:2008is}.
However, in experiments thermal fluctuations inevitably smear out the fine features in the density distribution. We thus propose to use the maximum entropy method \cite{Jarrell:1996uo} to extract the DOS from the noisy finite temperature density distributions. %The algorithm is shown to be stable against noise in the measurement and provide a way to resolve of the Hofstadter's fractal energy spectrum at achievable temperatures. 
Our simulations show that one is able to recover Hofstadter's butterfly solely from in situ imaging of the density profiles at current achievable temperature and resolution. 

\begin{figure*}[t]
\includegraphics[width=5.5cm]{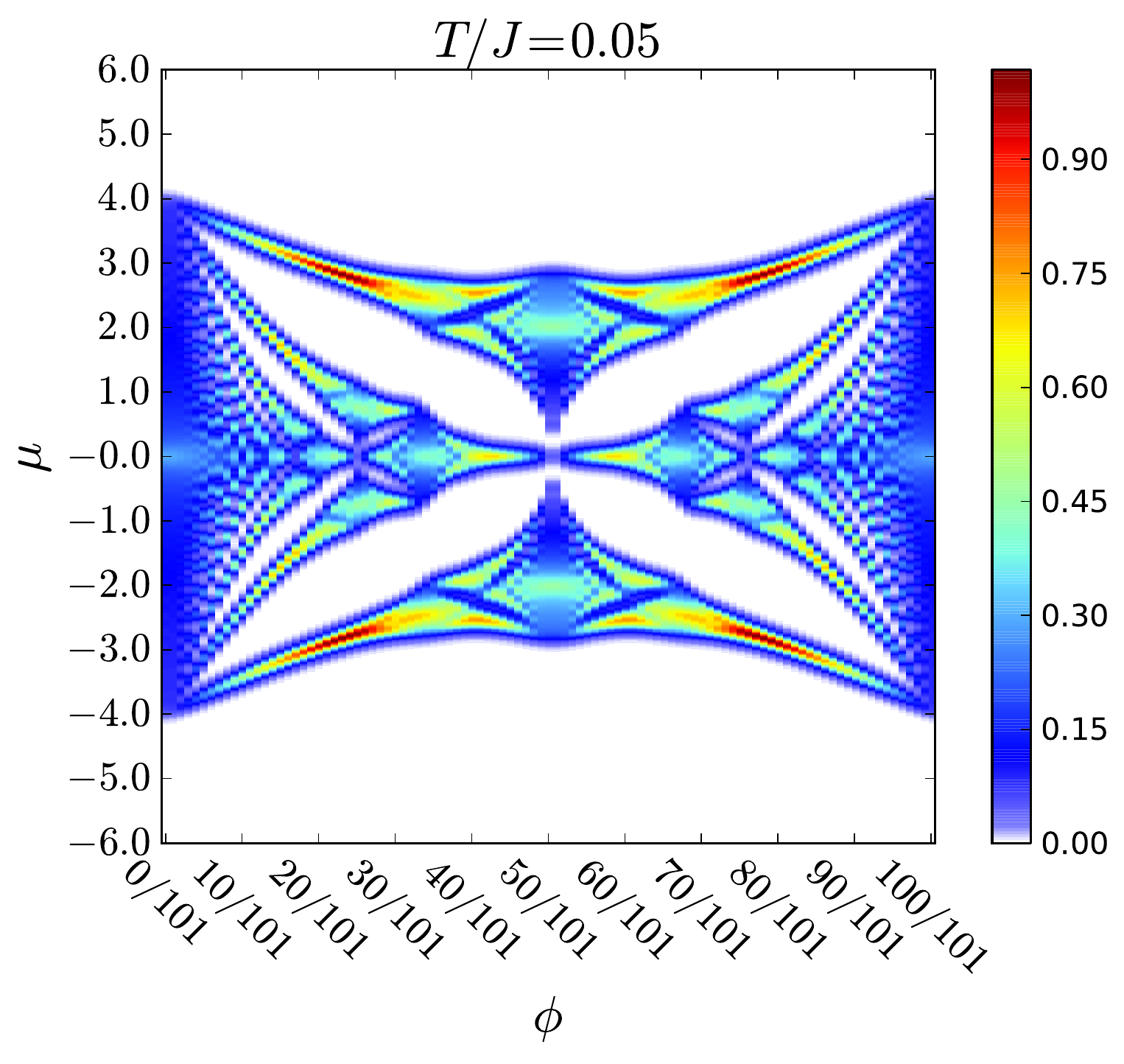}
\includegraphics[width=5.5cm]{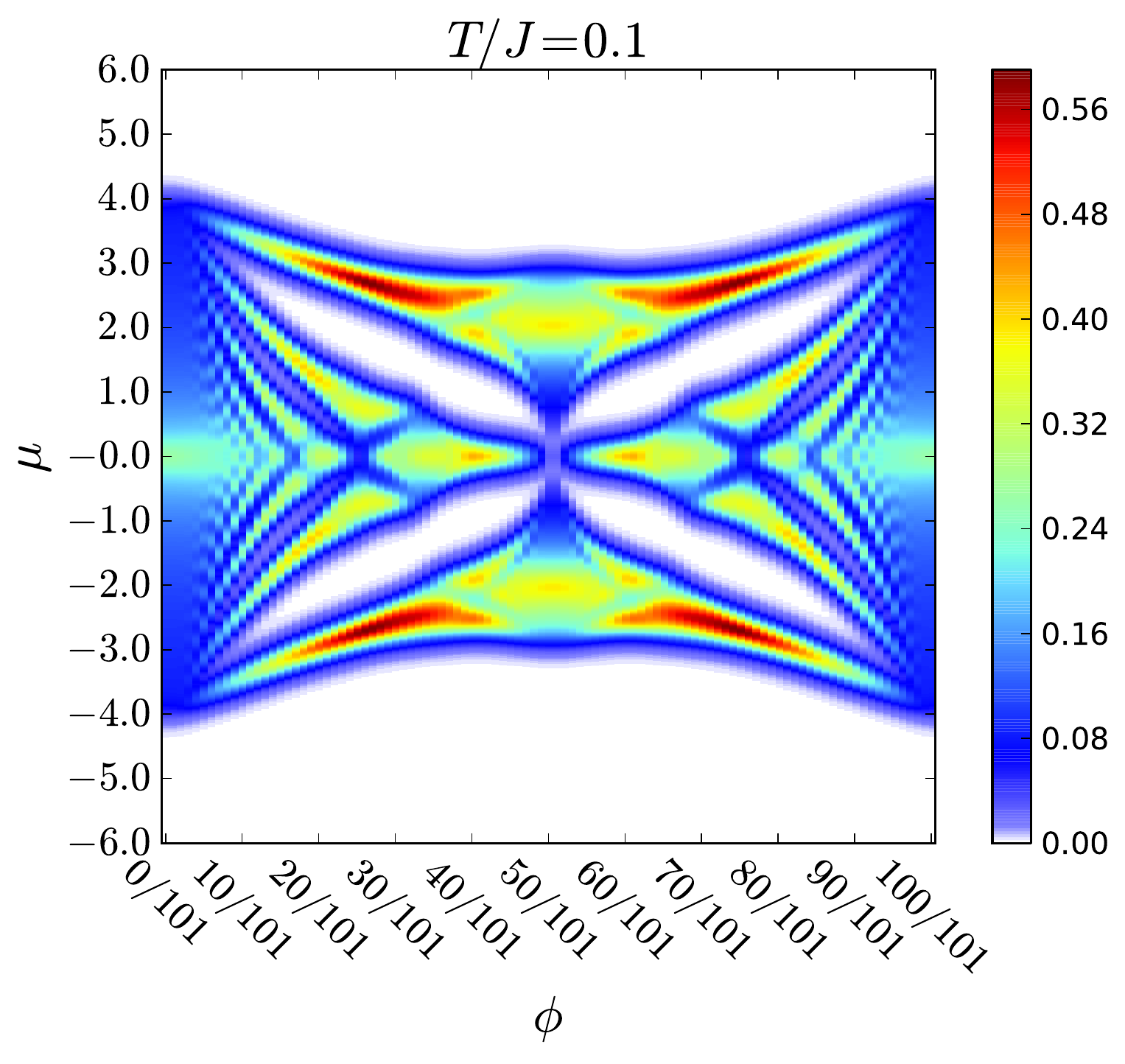}
\includegraphics[width=5.5cm]{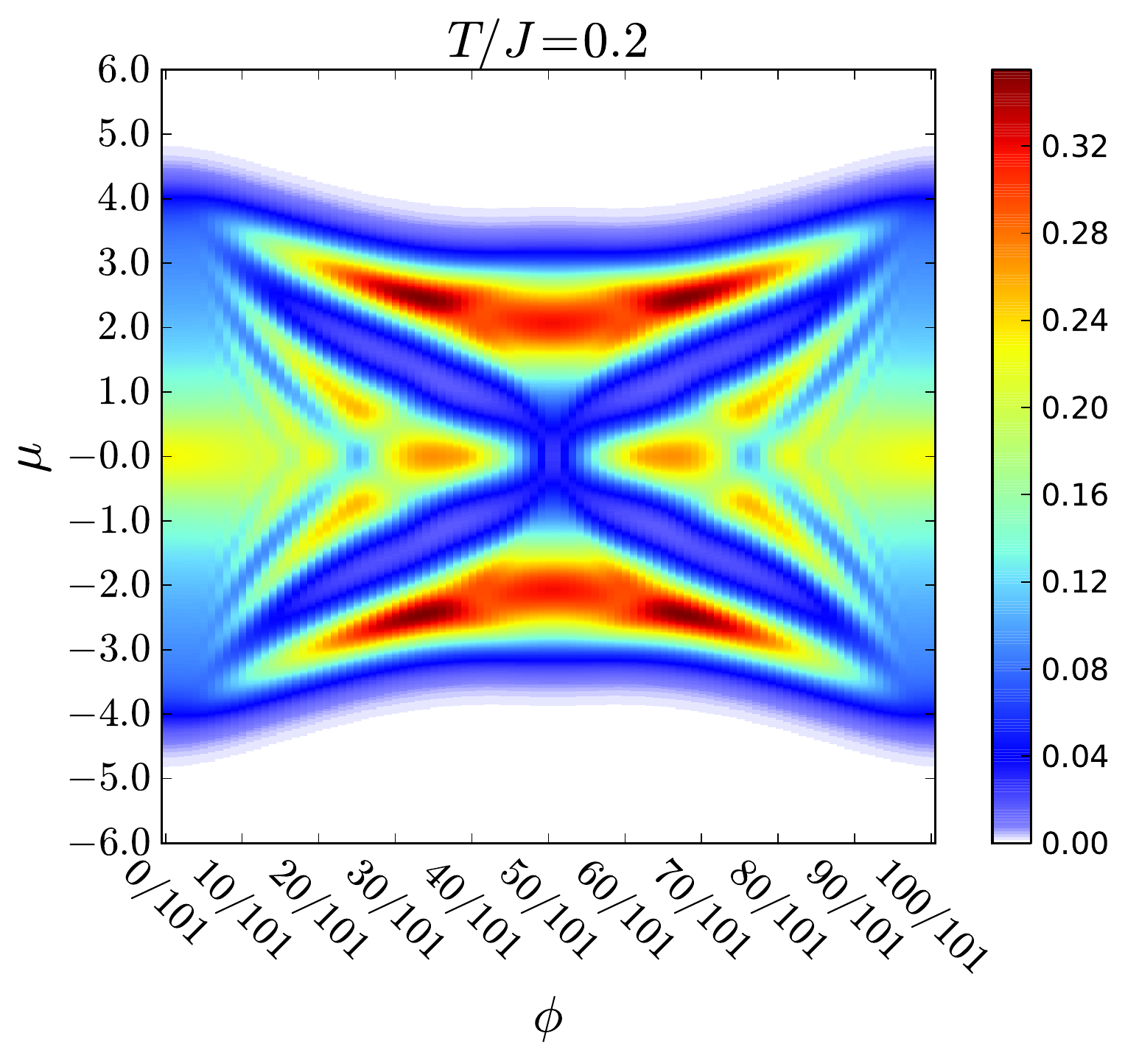}
\includegraphics[width=5.5cm]{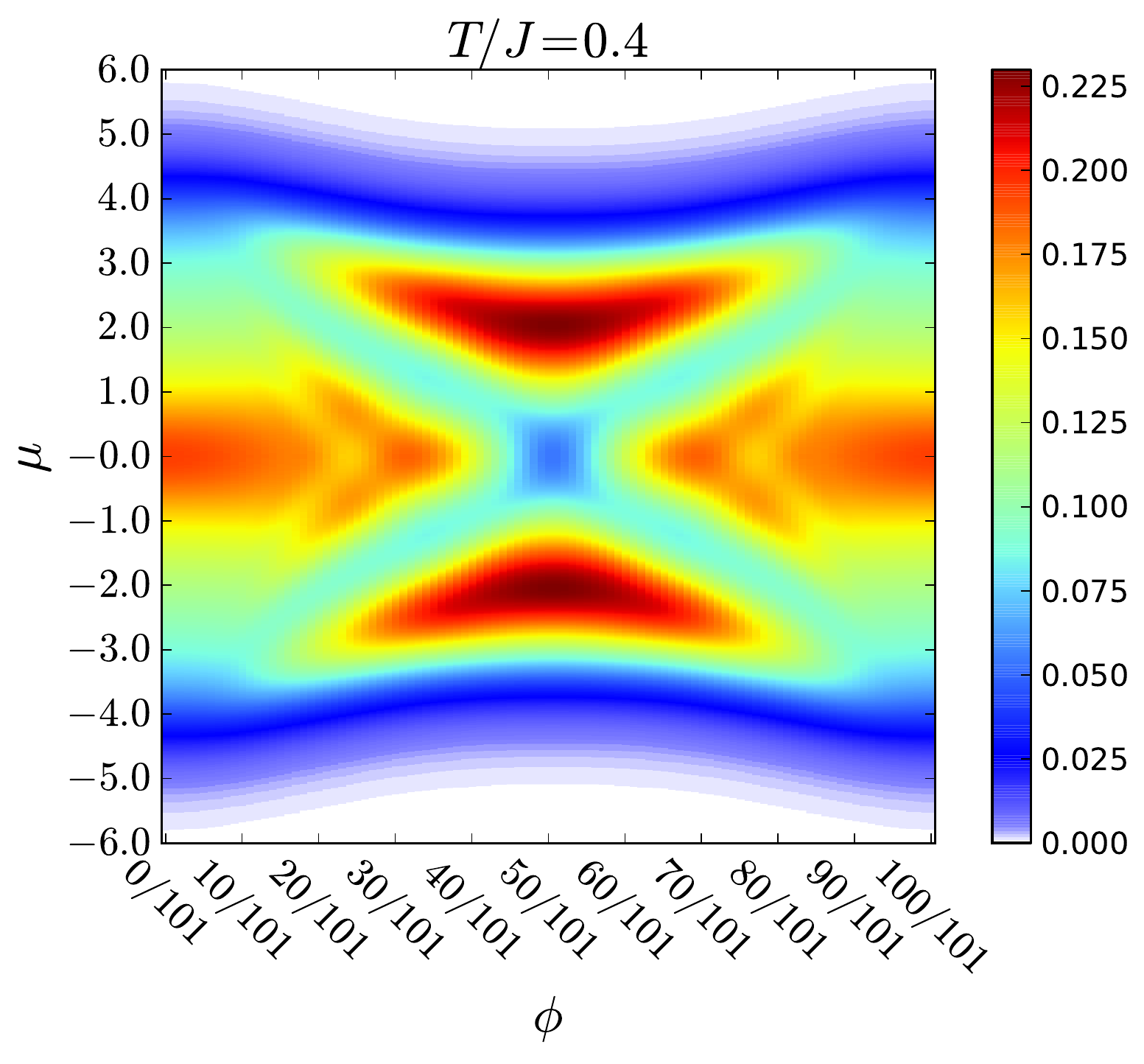}
\includegraphics[width=5.5cm]{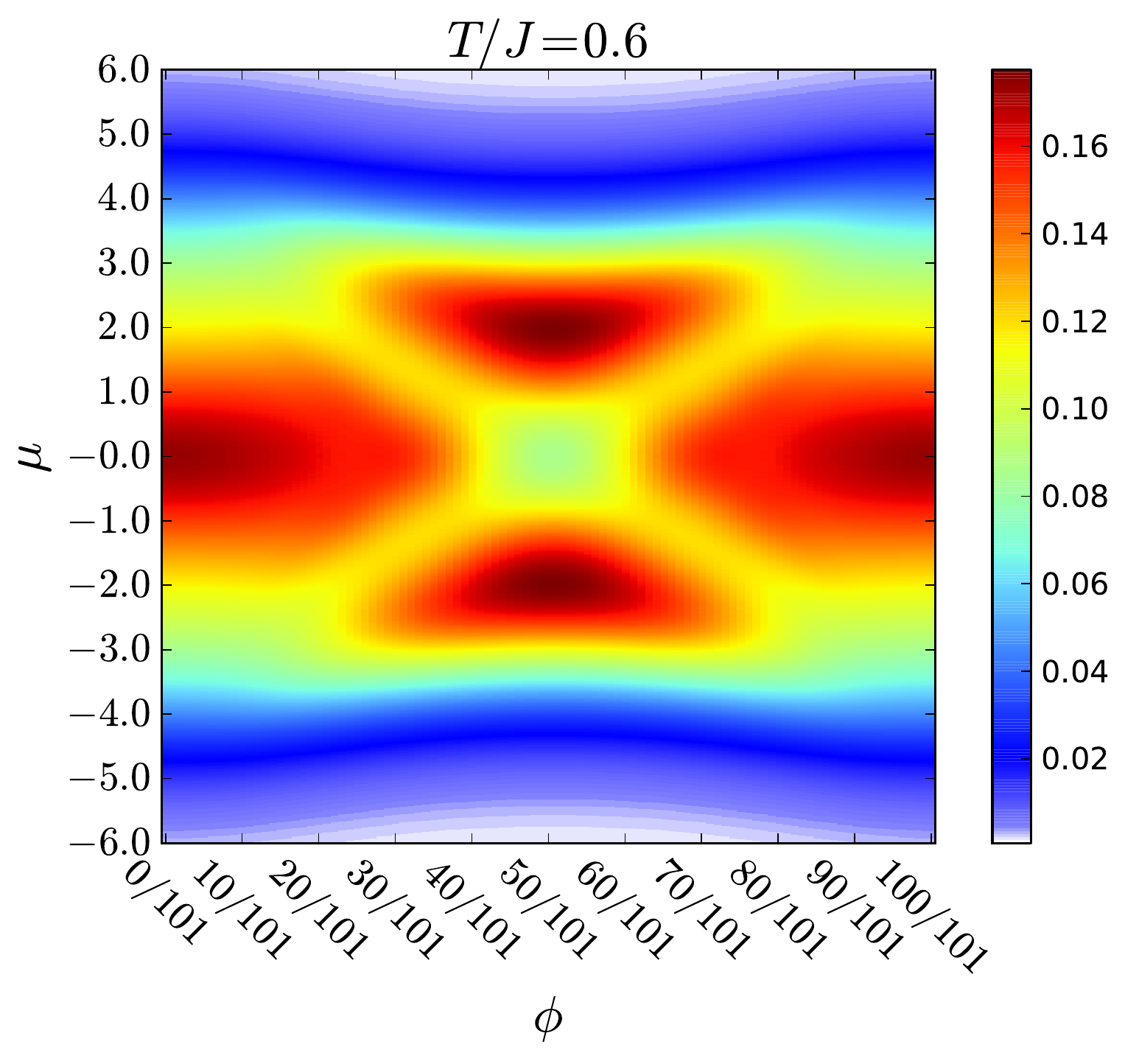}
\includegraphics[width=5.5cm]{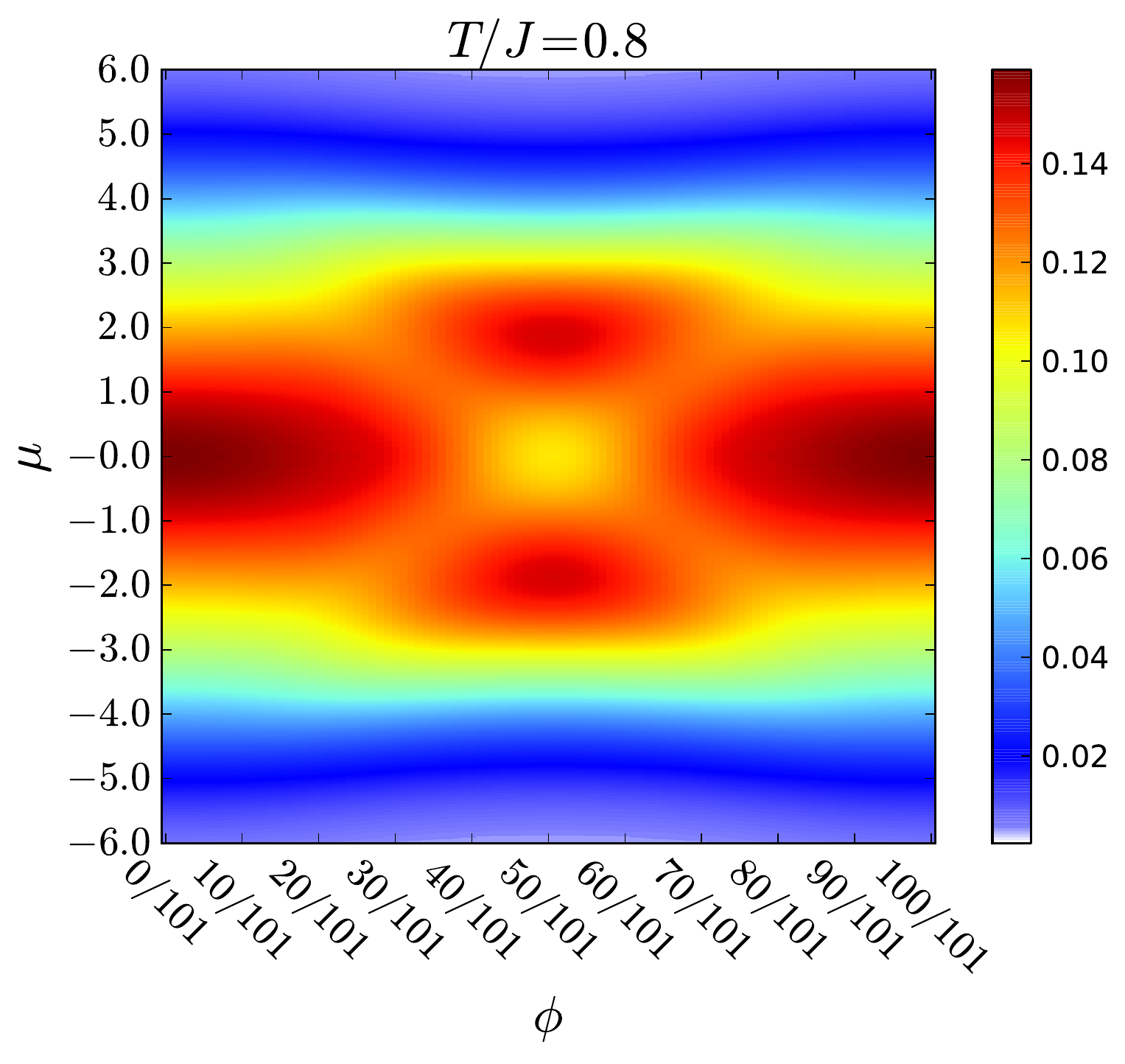}
\caption{The compressibility of the Hofstadter model versus the chemical potential $\mu$ and flux $\phi$ at different temperatures.}
\label{fig:EOS}
\end{figure*}

%{\bf The Model and Method}
The Hamiltonian of the Hofstadter model reads, 

\begin{equation}
H = -J \sum_{m,n} e ^{-i2\pi n\phi} \hat{c}^{\dagger}_{m+1,n} \hat{c}_{m,n}+\hat{c}^{\dagger}_{m,n+1} \hat{c}_{m,n} + H.c, 
\label{eq:Ham}
\end{equation}
where $J$ is the hopping amplitude and $\hat{c}_{m,n}$ are the fermionic annihilation operator, with $m$ and $n$ being the column and row indices of a square lattice. An atom hopping clock wise around a plaquette on the square lattice accumulates a phase $\phi$. Since the typical temperature in the optical lattice is higher than the energy scales associated with the fractal energy spectrum, it is essential to consider the finite temperature properties of the model. To calculate the thermodynamic properties, we adopt the exact diagonalization (ED)~\cite{Hasegawa:1989wq} and the quantum transfer matrix method (QTM)~\cite{Xu:2008gq,Yang:2012gf}. In both methods we choose $\phi=p/q$ where $p,q$ are two relatively prime integers. In the ED calculation, we diagonalize the Bloch Hamiltonian for each momentum and then calculate the thermodynamical quantities from the exact energy spectrum. In the QTM approach, we calculate the partition function of a system with fixed width and let the length grow to infinite. All other thermodynamic quantities can then be calculated from numerical differentiation of the grand-canonical thermodynamical potential. We have cross checked the results from both methods. 

The key physical observable is the density versus chemical potential, which is related to the DOS $D(\varepsilon)$ through \begin{equation}
 \rho(\mu,T) = \int^{\infty}_{-\infty} f\left(\frac{\varepsilon -\mu}{k_{B}T}\right) D(\varepsilon) \,\mathrm{d}\varepsilon,
 \label{eq:density}
\end{equation}
where $f(x)=1/(e^{x}+1)$ is the Fermi-Dirac distribution, $T$ is  temperature of the system. 
To probe the DOS, we take a derivative of both sides and get the compressibility
\begin{equation}
\kappa (\mu,T) \equiv\frac{\partial \rho}{\partial \mu} = \int^{\infty}_{-\infty} \frac{\partial f}{\partial \mu} D(\varepsilon) \,\mathrm{d}\varepsilon. 
\label{eq:compressibility}
\end{equation}
Since $\partial f/\partial \mu=f(1-f)/(k_{B}T)$ approaches to the Dirac delta function $\delta(\varepsilon-\mu)$ at zero temperature, the zero temperature compressibility directly probes the DOS~\footnote{Other thermodynamic quantities like entropy and specific heat also reveal the DOS in the zero temperature limit, however unlike the compressibility, they are not directly accessible in the cold atom toolbox.}:

\begin{equation}
D(\varepsilon) = \lim_{T\rightarrow0}\kappa(\varepsilon,T).
\label{eq:zeroTcompressibility}
\end{equation}

%\begin{equation}
%s(\mu)=-\int^{\infty}_{-\infty}[f\ln(f)+(1-f)\ln(1-f) ]D(\varepsilon) d\varepsilon
%\end{equation}
%approaches to $\frac{\pi^{2}T}{3} D(\mu)$ 

Figure~\ref{fig:EOS} shows the compressibility versus chemical potential and magnetic flux at different temperatures. At $T/J=0.05$ one can clearly see the fractal shape of the energy spectrum. The compressibility is zero when the chemical potential is in the energy gap. At higher temperature the fine features in the compressibility are smeared out, but the coarse feature of the butterfly remains. Even at $T/J=0.8$, the suppression of the compressibility close to $\mu=0,\phi=1/2$ is still visible. There the system has Dirac like dispersion around $\mu=0$ and the DOS vanishes linearly. This is in contrast to the $\phi=0$ case where the compressibility peaks at $\mu=0$ because of the Van Hove singularity in the DOS. %The suppression the compressibility and produces a plateau features in the denisty~\cite{Zhu:2007ja} like a gapful state. %The plot of entropy and specific heat shows similar features. 

%{\bf Local Compressibility Measurement}
In cold atom experiments the trapping potential provides a scan of the chemical potential which can be used to determine the density of states of a uniform system~\footnote{We need the local chemical potential in the trap to scan through the energy band, i.e. the density in the trap center to reach one. However, one can  utilize the particle-hole symmetry of the Hofstadter lattice and only probe $\rho<0.5$ to restore the whole spectrum.}. To see this we first treat the trapping potential using a local density approximation (LDA). The local chemical potential varies as $\mu(r)= \mu_{0}-\alpha r^{2}$ where $\mu_{0}$ is the chemical potential in the trap center, $\alpha$ is related to the geometric mean of the trapping frequencies and the atom mass, $r$ is the rescaled distances of a site to the trap center. The density $\rho(r)$ can be measured from the in situ imaging of the atomic cloud~\cite{Gemelke:2009ja, VanHoucke:2012ic,Ku:2012gra,Lee:2012jm}. The local compressibility can then be estimated as 
\begin{equation}
\kappa (r) =  -\frac{1}{2\alpha r}\frac{d\rho}{dr}.
\end{equation}
Combining this with the known $\mu(r)$ one can recover $\kappa(\mu)$ up to an overall shift of the chemical potential. Collecting measurements for different $\phi$, one can then recover the compressibility plots shown in Fig.~\ref{fig:EOS}~\footnote{$\mu_{0}$ and $\alpha$ will vary in different experimental runs. One can nevertheless utilize the particle-hole symmetry of the Hofstadter model and chose the origin of chemical potential at $\rho=0.5$ for each measurement.}.

\begin{figure}[t]
\includegraphics[width=8cm]{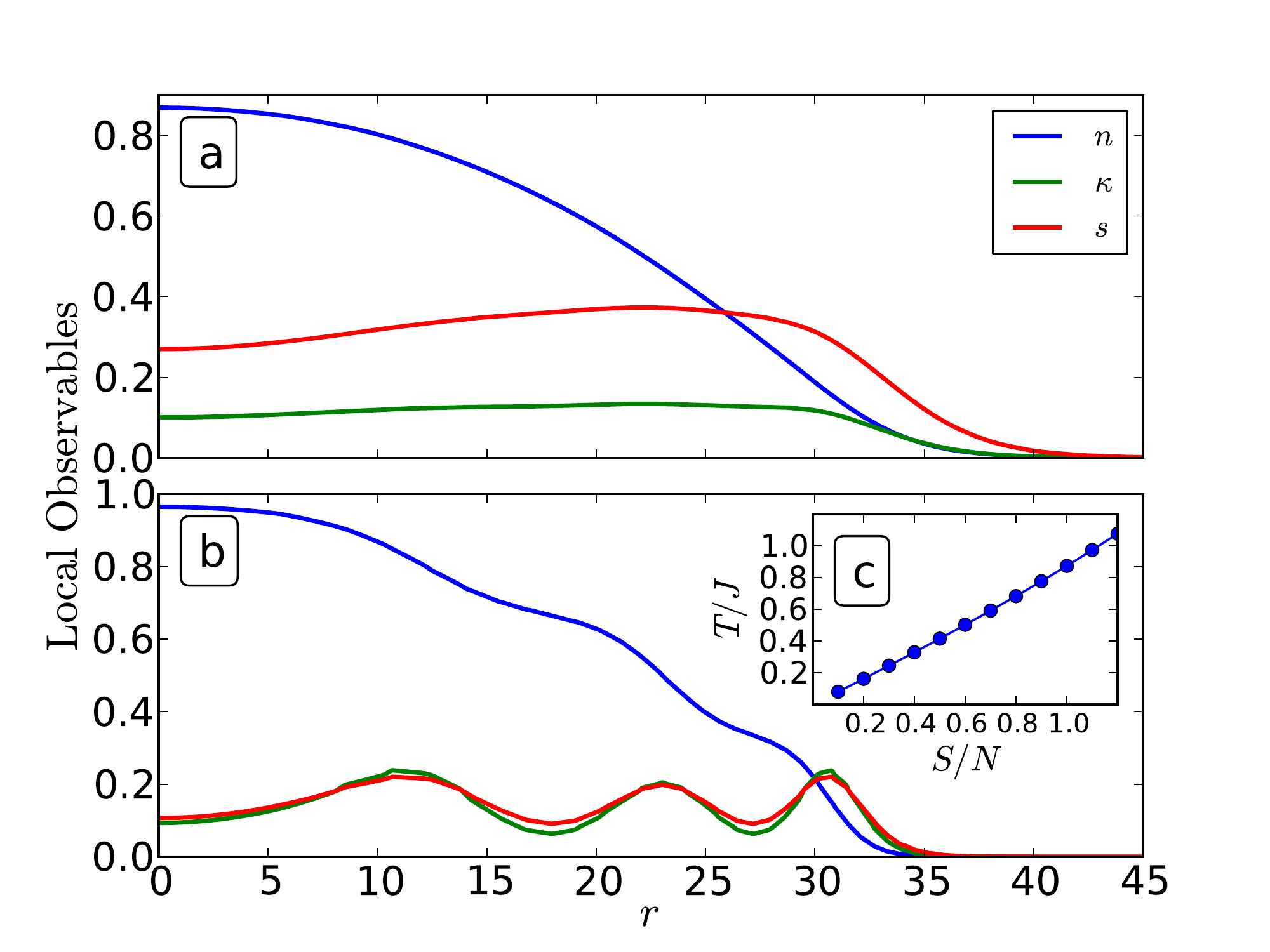}
\caption{{\bf Local observables in a 3D harmonic trap.} Solid lines show the density $\rho$, compressibility $\kappa$ and entropy $s$ with $N=60000,\alpha=0.006$ and $\phi=1/3$. The entropy per particle is (a) $S/N=1.0k_{B}$ and (b) $S/N=0.4k_{B}$. The corresponding temperatures is $T/J=0.873$ and $T/J=0.329$ respectively. (c). The temperature of the cloud versus entropy per particle $S/N$ calculated using LDA with the equation of state given by the quantum transfer matrix method. %{\bf [MT: Lei, how was this calculated?]} 
}
\label{fig:distribution}
\end{figure}

%Difficulties 
Finite temperature effects and sampling noise inevitably smear out the fine features in the density profile in experimental measurements. %The thermal fluctuations of the atom number increases with temperature~\cite{Zhou:2011dj,Sanner:2010hq,Muller:2010ei} according the fluctuation-dissipation theorem (FDT) 
Noisy signals pose problems for extracting the local compressibility from the density distributions, which raises the question whether is it possible to observe the fractal structure of $\kappa$ at an experimental accessible temperature. %Since current experiment has more control over the total entropy, we study thermal distribution in a 3D cloud with  fixed entropy per particle in following. 
%Estimate temperatures 
Figure~\ref{fig:distribution}(a) shows results for $N=60000$ fermions in a three-dimensional (3D) trap with $\alpha=0.006,\phi=1/3$ and entropy per particle $S/N= 1.0k_{B}$, which is currently easily accessible~\cite{McKay:2011fc}. It corresponds to a temperature $T/J=0.873$ and the fine features in local observables has been smeared out. %Ref.~\cite{Umucallar:2008fq} reported similar results with the LDA calculation in a 2D trap. 
Only when the entropy per particle is reduced to $S/N=0.4k_{B}$ (Fig~\ref{fig:distribution}(b)), one can directly observe the density plateaus and the corresponding peaks in the local compressibility. To resolve a particular feature in the energy spectrum requires the temperature to be smaller than the corresponding energy gap. Figure~\ref{fig:distribution}(c) shows the temperature in unit of $J$ versus the entropy per particle. The required entropy for resolving the plateau at $\phi=1/3$ is comparable to achieving the antiferromagnetic states in the 3D Hubbard model~\cite{Fuchs:2011cha}, which is already a challenging task. Above analysis show that even without any noise, the thermal broadening effect already makes it difficult to resolve the Hofstadter butterfly from the local compressibility at experimentally achievable temperatures. 

%%Anisotropic hopping 
%Finally, as in the experiment realizes Hofstadter model with anisotropic hoppings where the ratio of the hopping amplitudes could be tuned with the intensity of the Raman laser. It is natural to ask whether the anisotropy helps detecting the fractal energy spectrum, given the average entropy per particle fixed. As this is the case observed in the experiments in anisotropic Hubbard model, where the anisotropy greatly enhence the n.n. spin correlations measured in that experiment. Fig.~\ref{eq:distribution} shows the density for anisotropic Hofstadter model with $J_{x}/J_{y}=1/2$. As shown there, the density plateau is less pronounces in the anisotropic with fixed entropy per particle. As we conclude to observe the fractal energy spectrum, it is advised to work with the Hofstadter model with isotropic hopping. 

%{\bf Maximum Entropy Spectral Analysis}
%Fit or smooth the density data wont't solve the problem, as one wills still suffer from the the thermal broadening effect. Here we want to introduce a solution to deal with the broadening and noise at the same time. 
We now come to the key idea of this paper: one is nevertheless able to restore the density of states $D(\varepsilon)$ from a seemingly featureless and noisy thermal density distribution using  techniques of  spectral analysis. Knowing the temperature of the system (which we will discuss in the following), one can directly deconvolute the effect of the Fermi-Dirac distribution in the Eq.(\ref{eq:density}) or Eq.(\ref{eq:compressibility}) to get $D(\varepsilon)$ from the density distributions. The zero temperature compressibility detection discussed above is a limiting case where one trivially deconvolutes a Dirac delta function in Eq.(\ref{eq:compressibility}). %One could also cast the deconvolution problem into a linear algebra equation, treating the Fermi-Dirac function an integral kernel. However, as temperature increases the kernel is singular and both approaches face difficulties. 

At high temperature it is in general difficult to deconvolute  Eq.(\ref{eq:density}) as it is an ill-posed problem, especially given the experimental uncertainties in the measured equation of state $\rho(\mu)$. To solve the difficulty, the maximum entropy method~\cite{Jarrell:1996uo} treats $D(\varepsilon)$ as a probability distribution and searches for the best solution (in the Bayesian sense) that is consistent with the measured data. The stochastic inference approach~\cite{Sandvik:1998ut,Mishchenko:2000vm,Beach:2004uc,Fuchs:2010it} employs a stochastic process and represents the resulting spectrum as an ensemble average of many feasible solutions. Recently, a new method based on the consistent constraints was also been proposed~\cite{Boris}. These methods have been used for the analytical continuation from the imaginary time quantum Monte Carlo data to the real frequency spectral functions~\cite{Silver1996,Jarrell:1996uo}. The deconvolution of Eq.(\ref{eq:density}) is related to the analytical continuation by setting the imaginary time  $\tau=0^{-}$ and introduce chemical potential dependence to the imaginary time Green's function. 

\begin{figure}[t]
\includegraphics[width=8cm]{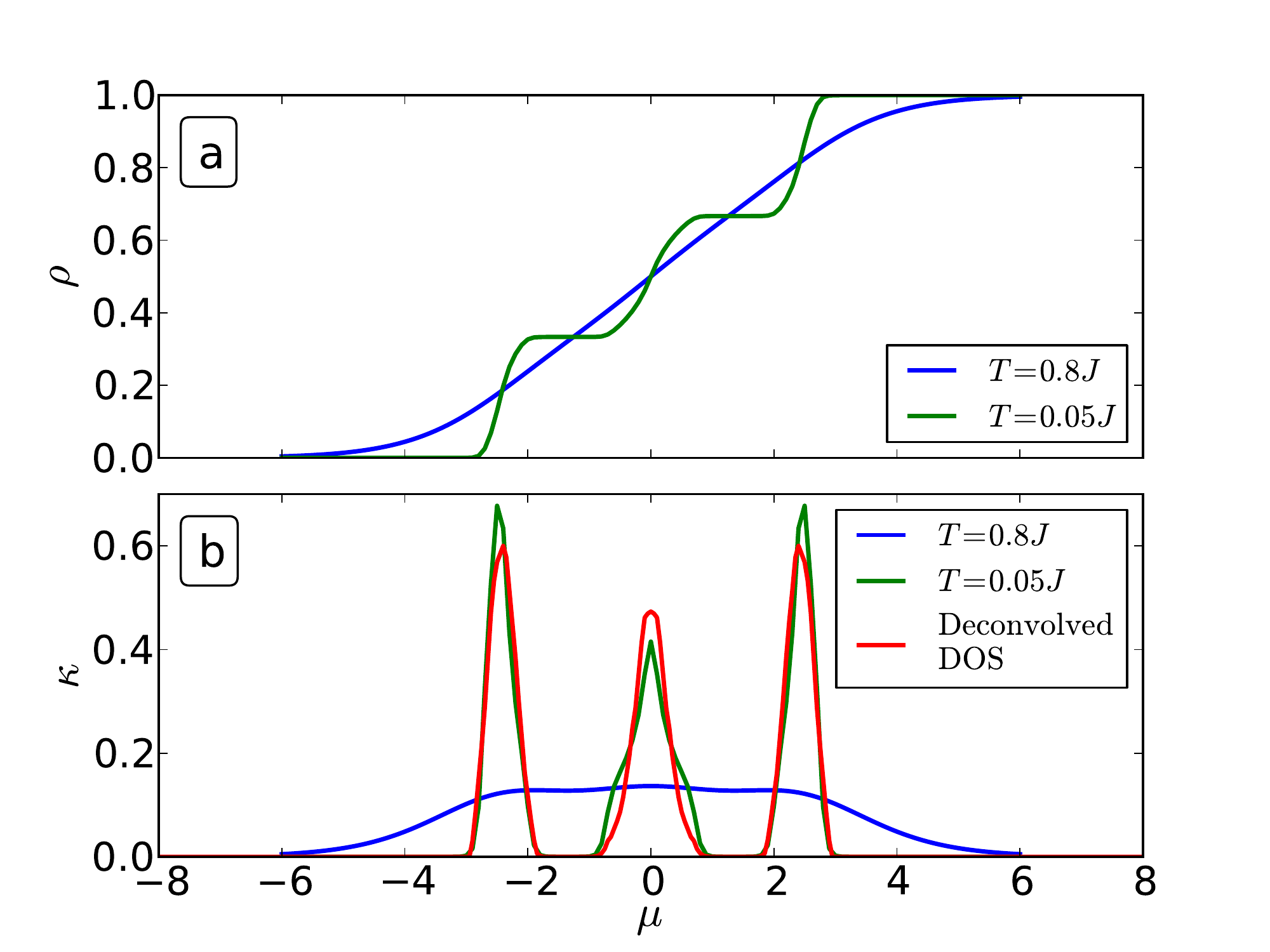}
\caption{{\bf Deconvolve the thermal broadening effect.} (a). Density $\rho$ versus chemical potential $\mu$ at high ($T=0.8J$) and low ($T=0.05J$) temperatures, $\phi=1/3$. (b). The corresponding compressibilities  which at low temperature ($T=0.05J$) reveals the DOS of the system. The red line shows the DOS restored from the $T=0.8J$ density distribution using the maximum entropy algorithm.}
\label{fig:MEM}
\end{figure}

We first apply the maximum entropy approach~\footnote{
For the spectral analysis calculation, we used the stochastic inference approach as implemented in~\cite{Sandvik:1998ut} and the maximum entropy code implemented in the ALPS package~\cite{BBauer:2011tz}. We have enforced the particle-hole symmetry in the resulting DOS.
} to a noiseless high temperature density distributions and show it is able to deconvolute the thermal broadening effect. Figure~\ref{fig:MEM}(a) shows $\rho(\mu)$ at $T/J=0.8$ which seems featureless compared to the density at $T/J=0.05$. Figure~\ref{fig:MEM}(b) shows the corresponding compressibilities, where the one at $T/J=0.05$ approximates the exact DOS well while the $T/J=0.8$ one is much broader. Neverthless, the deconvoluted DOS from the density at $T/J=0.8$ agrees well with $\kappa(T=0.05J)$, Figure~\ref{fig:MEM}(b). In particular, from the seemingly featureless density profile at high temperature, we have restored the three peaks in the DOS, corresponding to the three energy bands at $\phi=1/3$. %For $\phi=1/2$ the deconvolved DOS also correctly reveals the vanishing DOS at $\varepsilon = 0$ originate from the Dirac band touching point of the Hofstadter model with even $q$. 
%Therefore the maxent is able to deconvole the thermal broderning effect from high temperature noiseless densities. 
%stable against noises 

\begin{figure}[t]
\includegraphics[width=8cm]{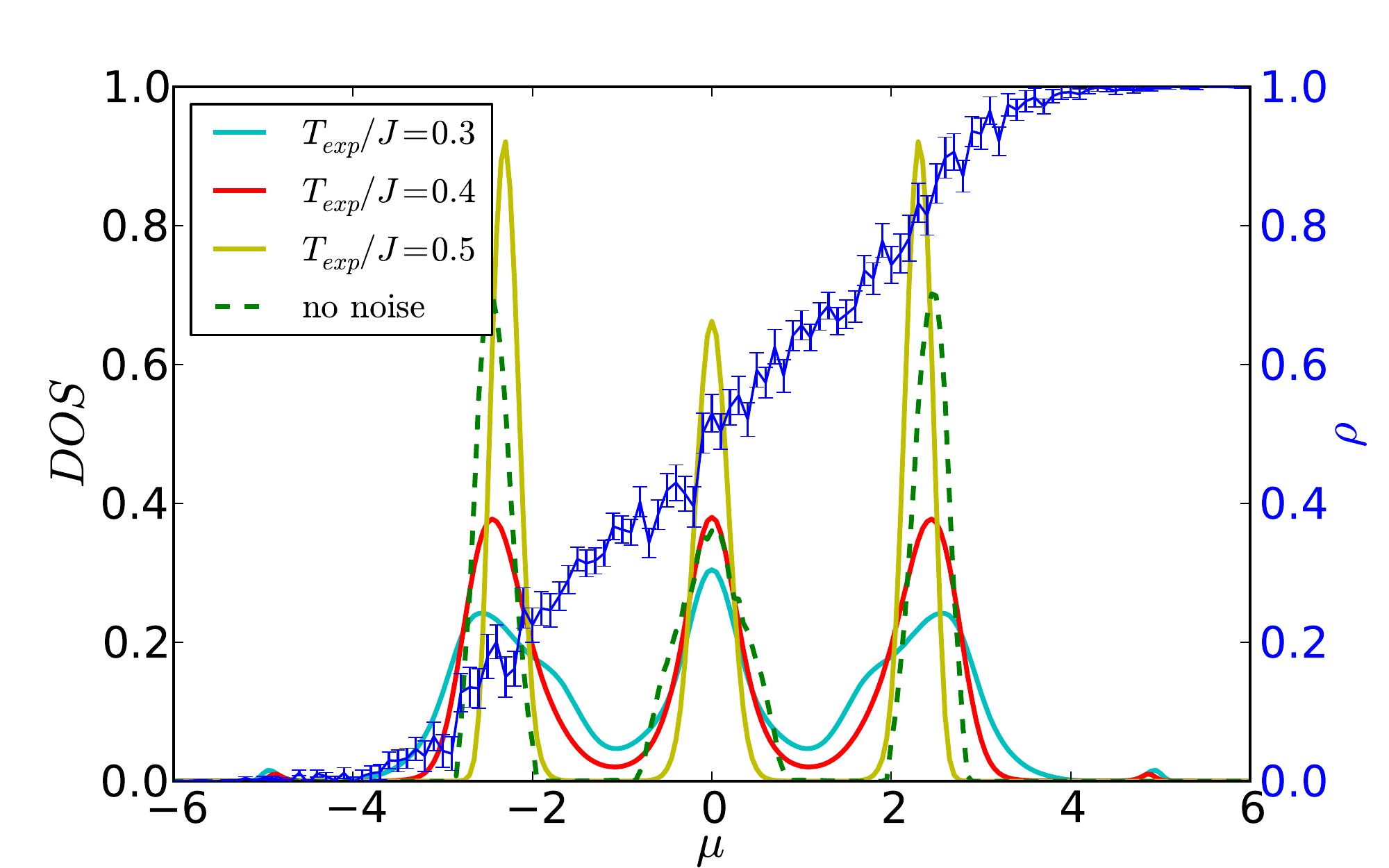}
\caption{{\bf Restoring the DOS from noisy data and imperfect temperature measurement.} The blue line with errorbars show the density profile at $T/J=0.4$ with noises following the fluctuation-dissipation theorem. The three solid lines (cyan, red and yellow) show DOS restored from the noisy data using temperature $T_{exp}/J=0.3,0.4$ and $0.5$ in the Fermi-Dirac kernel. The dashed green line shows the DOS restored from the exact $\rho(\mu)$ data.}
\label{fig:noisyMEM}
\end{figure}

To mimic noisy experimental measurements, we generate Gaussian distributed random numbers with standard deviation $\sqrt{\kappa T}$ according to the fluctuation-dissipation theorem (FDT) ~\cite{Zhou:2011dj,Sanner:2010hq,Muller:2010ei} and add them to the exact $\rho(\mu)$ data. We then feed the average values and statistical errors of $100$ noisy samples (Fig.\ref{fig:noisyMEM}) to the maximum entropy calculation. Random noise further washes out the fine features in the density and the resulting DOS are  broader than the exact one. (compare in Fig.\ref{fig:noisyMEM} the red solid line against the dashed green  line.) Nevertheless, the deconvoluted DOS based on the noisy data still correctly restores the three peaks correspond to the $\phi=1/3$. 

%determine Temp  
To determine the integral kernel $f(\frac{\varepsilon-\mu}{k_{B}T})$ one needs to determine the temperature $T$ of the system, which can be done using several approaches~\cite{McKay:2011fc}. In particular, when the density profile is available, one could fit the density in the wing of the cloud~\cite{Ho:2009cp} (with theoretical input about the $\rho(\mu)$ in the dilute limit) or using the FDT~\cite{Zhou:2011dj, Ma:2010do, Gemelke:2009ja} to determine the temperature.
%Effect of error in T to the MEM result, is it very sensitive ? 
We examine the effect of error in the measured temperature $T_{exp}$ on the restored DOS in Fig.\ref{fig:noisyMEM}. If $T_{exp}>T$, the deconvolution results in a sharper DOS and the peak position are shifted, while $T_{exp}<T$ has the opposite effect. Still, the error in the measured temperature $T_{exp}\neq T$ does not destroy the overall feature of the DOS. This analysis also shows that the deconvolution is stable against small variations of temperature in different experimental runs.

\begin{figure}[t]

\includegraphics[width=4.25cm]{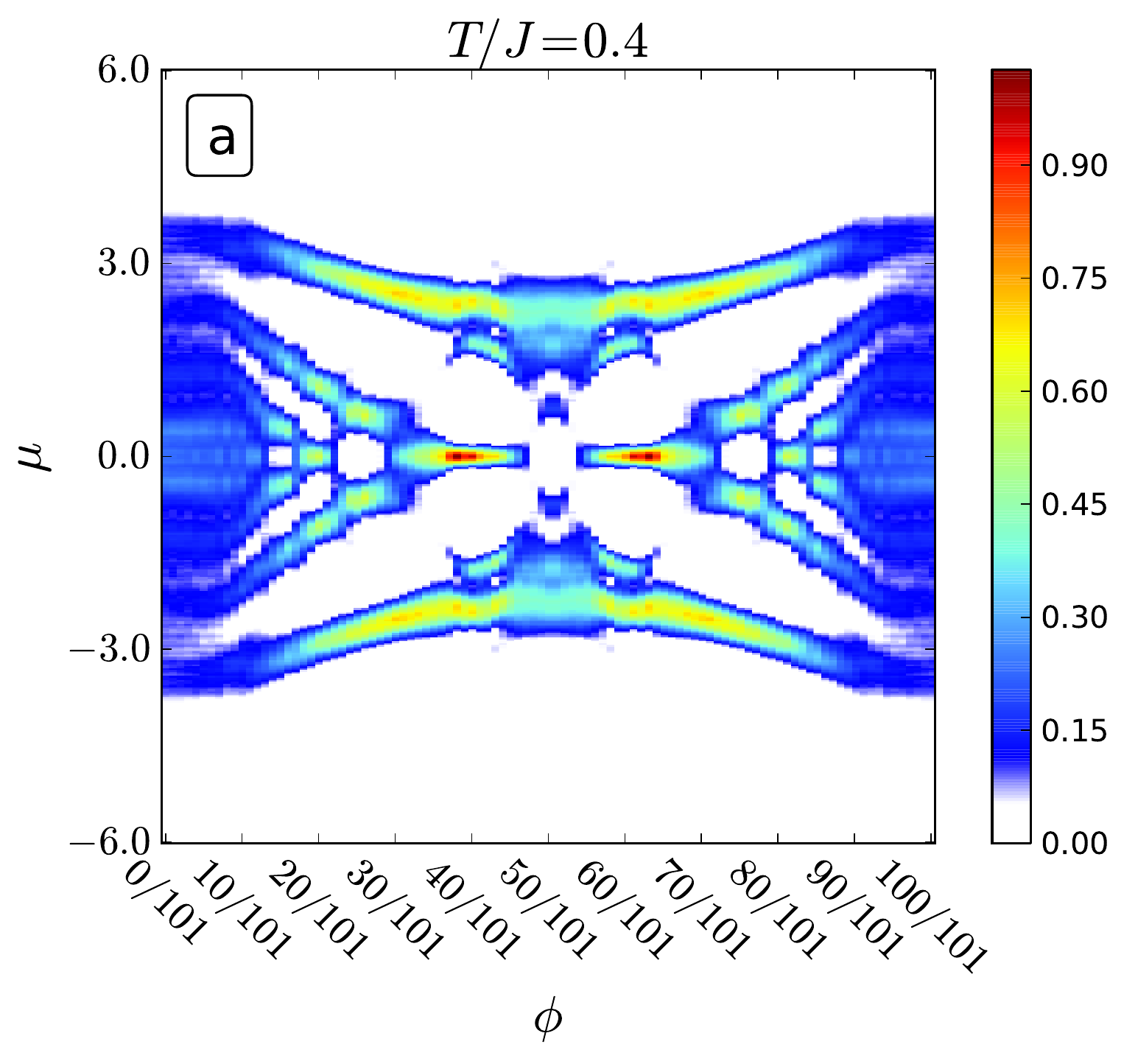}
\includegraphics[width=4.25cm]{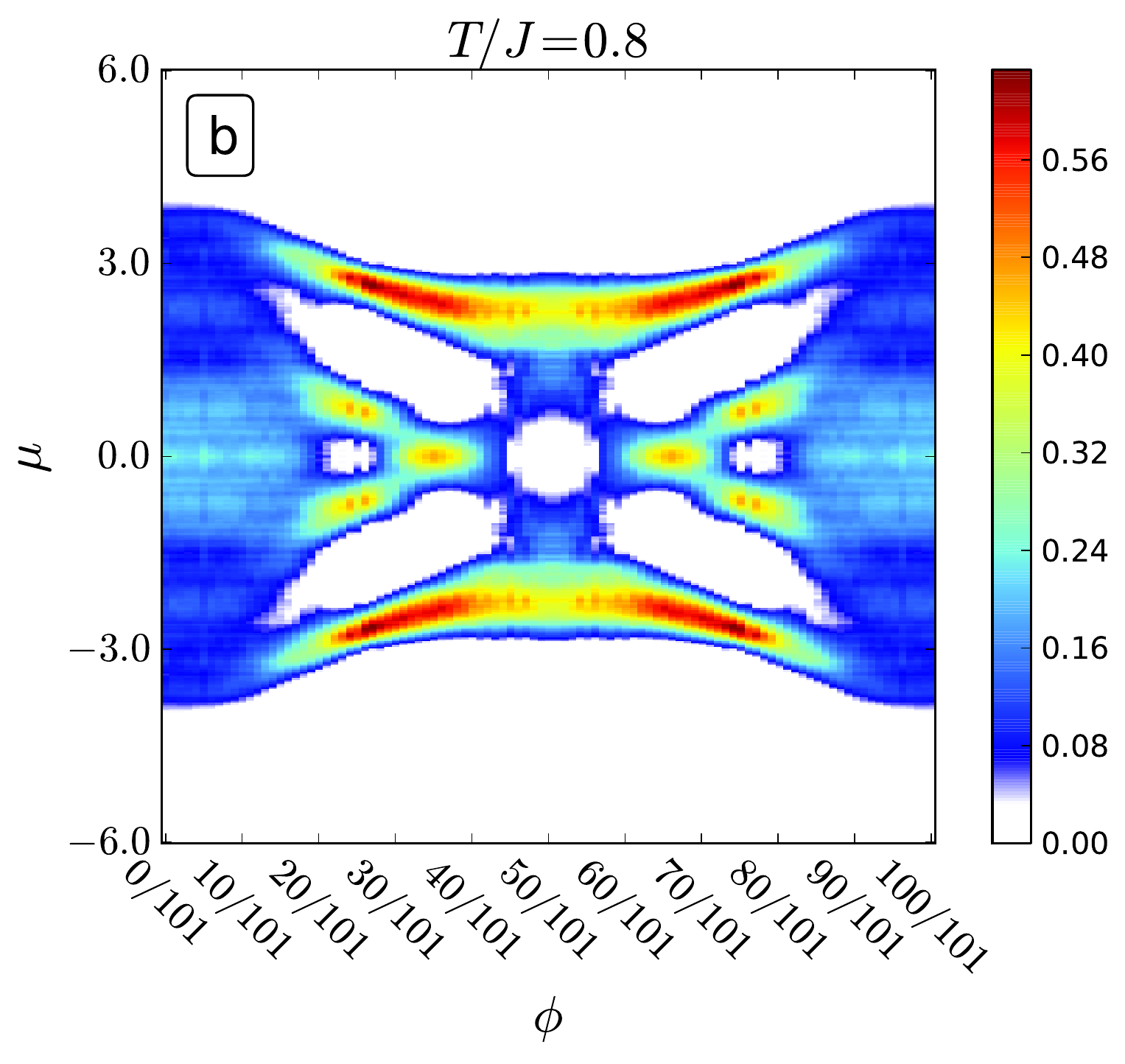}
\includegraphics[width=4.25cm]{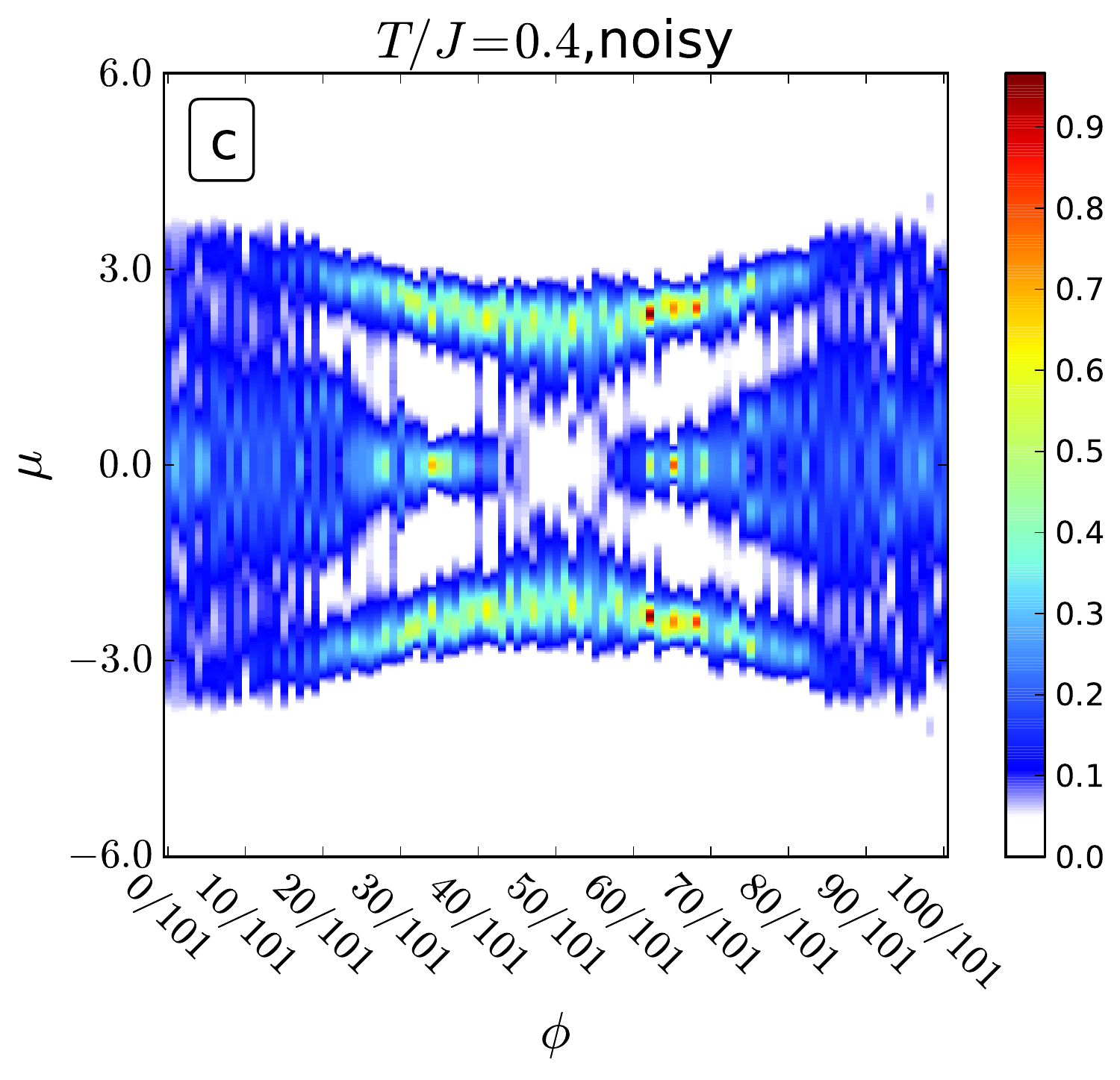}
\includegraphics[width=4.25cm]{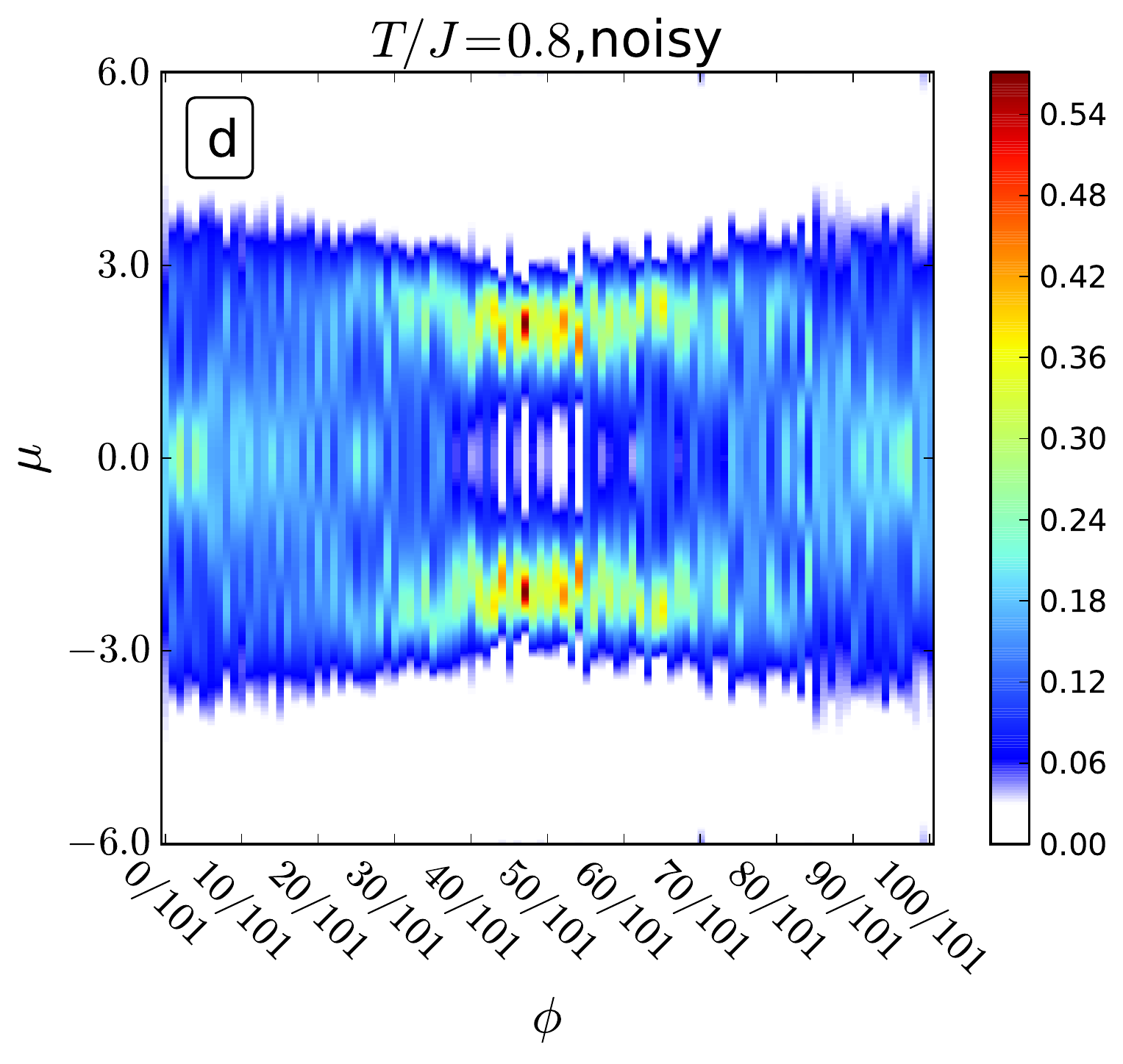}

\caption{{\bf The deconvoluted Hofstadter spectrum using  thermal density distributions.} (a-b) without noises (infinite samples). (c-d). obtained from $100$ noisy samples.}
\label{fig:MEMbutterfly}
\end{figure}

Finally, we show the deconvoluted DOS at $T/J=0.4$ and $0.8$ with different $\phi$ in the Fig.~\ref{fig:MEMbutterfly}(a-b). To further incorporate corrections beyond the local density approximation (LDA)~\footnote{The LDA correction to the density profile is less than $0.1\%$~\cite{Zhou:2011dj} at the typical experimental temperatures ($T/J\sim 1$). We have also performed the deconvolution using the LDA density profiles. The images are only slightly different from the one using the exact density data, indicating the the LDA corrections to the final results are indeed small. These corrections will further decreases if one uses a weaker trapping potential experimentally. }, we use exact densities on a $101^2$ lattice in a trapping potential with $\alpha=0.006$. The deconvoluted DOS reproduces the butterfly spectrum at both temperatures, although the fine structures around the edge ($\phi\sim0$ and $\phi\sim 1$) are smeared out. Figure~\ref{fig:MEMbutterfly}(c-d) shows the DOS restored from noisy density data. Even at high temperature $T/J=0.8$ one can still observe the reminiscent of the butterfly spectrum, where the suppression of $D(\mu=0)$ at $\phi=1/2$ compares to the $\phi=0$ case is the most significant feature. By further decreasing the temperature (Fig.~\ref{fig:MEMbutterfly}(c)), one can uncover finer structures of Hofstadter's butterfly. More importantly, our simulation shows that it is more favorable to gather better statistics, which already allows one to restore the core feature of Hofstadter's butterfly (Fig.~\ref{fig:MEMbutterfly}(b)) at high temperature ($T=0.8J$). This is encouraging news for experimentalists who want to observe the butterfly spectrum.

%{\bf Summary and Outlook}
In a broader context, our work provides an example of how spectral information can be extracted from static thermodynamic properties (the equation of state). In the context of this paper, a maximum entropy analysis of the noisy finite temperature density profiles turns out to yield sufficient information to allow observation of the Hofstadter butterfly in ultracold atomic Fermi gases. This approach can also be used in free space and other complicate optical lattices structures, to reveal the novel Dirac dispersions and flat bands~\cite{Tarruell:2012db, Jo:2012br}. A generalization of our method to interacting systems would be of great interest.

%Generalization to the free space or momentum resolved case may lead to further useful applications. 

%We propose to detect the Hofstadter's fractal energy spectrum from the density distributions of atomic Fermi gases. At low temperature the local compressibility directly probes the density of state of the system. To overcome the thermal broadening and noises in the measurements, we use the maxent method to directly extract DOS from noisy finite temperature density distributions. We show it is possible to reveal the overall features of the fractal spectrum in the realistic experimental situations with a current achievable entropy value. Further improvement of experimental techniques may allow to uncover the more intriguing features in the Hofstadter's butterfly.  

%Moreover, applying the proposed  approach to interacting systems allows one to extract the finite-temperature spectral function and address the pseudogap physics of strongly correlated fermionic systems~\cite{Ding1996, Gaebler:2010kw}. %Generalizing the approach to bosons may allow a  spectral analysis of superfluid-Mott transitions \cite{Greiner:2002wt}. 

{\bf Acknowledgment}
We thank Jakub Imri\v ska, Daniel Greif and Hiroshi Shinaoka for useful discussions. The work was supported by the Swiss National Science Foundation through the NCCR QSIT, the European Research Council, and the Aspen Center for Physics under National Science Foundation grant number 1066293. Simulations were performed on the Brutus cluster at ETH Zurich.

\bibliographystyle{apsrev4-1}
%\bibliography{/Users/wanglei/Documents/Papers2/papers}
\bibliography{ThermalHofstadter}

%merlin.mbs 2010-03-15 4.21a (PWD, AO, DPC)
%Control: key (0)
%Control: author (72) initials jnrlst
%Control: editor formatted (1) identically to author
%Control: production of article title (-1) disabled
%Control: page (0) single
%Control: year (1) truncated
%Control: production of eprint (0) enabled
\begin{thebibliography}{46}%
\makeatletter
\providecommand \@ifxundefined [1]{%
 \@ifx{#1\undefined}
}%
\providecommand \@ifnum [1]{%
 \ifnum #1\expandafter \@firstoftwo
 \else \expandafter \@secondoftwo
 \fi
}%
\providecommand \@ifx [1]{%
 \ifx #1\expandafter \@firstoftwo
 \else \expandafter \@secondoftwo
 \fi
}%
\providecommand \natexlab [1]{#1}%
\providecommand \enquote  [1]{``#1''}%
\providecommand \bibnamefont  [1]{#1}%
\providecommand \bibfnamefont [1]{#1}%
\providecommand \citenamefont [1]{#1}%
\providecommand \href@noop [0]{\@secondoftwo}%
\providecommand \href [0]{\begingroup \@sanitize@url \@href}%
\providecommand \@href[1]{\@@startlink{#1}\@@href}%
\providecommand \@@href[1]{\endgroup#1\@@endlink}%
\providecommand \@sanitize@url [0]{\catcode `\\12\catcode `\$12\catcode
  `\&12\catcode `\#12\catcode `\^12\catcode `\_12\catcode `\%12\relax}%
\providecommand \@@startlink[1]{}%
\providecommand \@@endlink[0]{}%
\providecommand \url  [0]{\begingroup\@sanitize@url \@url }%
\providecommand \@url [1]{\endgroup\@href {#1}{\urlprefix }}%
\providecommand \urlprefix  [0]{URL }%
\providecommand \Eprint [0]{\href }%
\@ifxundefined \urlstyle {%
  \providecommand \doi  [0]{\begingroup \@sanitize@url \@doi}%
  \providecommand \@doi [1]{\endgroup \@@startlink {\doibase
  #1}doi:\discretionary {}{}{}#1\@@endlink }%
}{%
  \providecommand \doi  [0]{doi:\discretionary{}{}{}\begingroup
  \urlstyle{rm}\Url }%
}%
\providecommand \doibase [0]{http://dx.doi.org/}%
\providecommand \Doi [0]{\begingroup \@sanitize@url \@Doi }%
\providecommand \@Doi  [1]{\endgroup\@@startlink{\doibase#1}\@@Doi}%
\providecommand \@@Doi [1]{#1\@@endlink}%
\providecommand \selectlanguage [0]{\@gobble}%
\providecommand \bibinfo  [0]{\@secondoftwo}%
\providecommand \bibfield  [0]{\@secondoftwo}%
\providecommand \translation [1]{[#1]}%
\providecommand \BibitemOpen [0]{}%
\providecommand \bibitemStop [0]{}%
\providecommand \bibitemNoStop [0]{.\EOS\space}%
\providecommand \EOS [0]{\spacefactor3000\relax}%
\providecommand \BibitemShut  [1]{\csname bibitem#1\endcsname}%
%</preamble>
\bibitem [{\citenamefont {Hofstadter}(1976)}]{Hofstadter:1976wt}%
  \BibitemOpen
  \bibfield  {author} {\bibinfo {author} {\bibfnamefont {D.~R.}\ \bibnamefont
  {Hofstadter}},\ }\href {http://prb.aps.org/abstract/PRB/v14/i6/p2239_1}
  {\bibfield  {journal} {\bibinfo  {journal} {Phys. Rev. B},\ }\textbf
  {\bibinfo {volume} {14}},\ \bibinfo {pages} {2239} (\bibinfo {year}
  {1976})}\BibitemShut {NoStop}%
\bibitem [{\citenamefont {Albrecht}\ \emph {et~al.}(2001)\citenamefont
  {Albrecht}, \citenamefont {Smet}, \citenamefont {von Klitzing}, \citenamefont
  {Weiss}, \citenamefont {Umansky},\ and\ \citenamefont
  {Schweizer}}]{Albrecht:2001kz}%
  \BibitemOpen
  \bibfield  {author} {\bibinfo {author} {\bibfnamefont {C.}~\bibnamefont
  {Albrecht}}, \bibinfo {author} {\bibfnamefont {J.}~\bibnamefont {Smet}},
  \bibinfo {author} {\bibfnamefont {K.}~\bibnamefont {von Klitzing}}, \bibinfo
  {author} {\bibfnamefont {D.}~\bibnamefont {Weiss}}, \bibinfo {author}
  {\bibfnamefont {V.}~\bibnamefont {Umansky}}, \ and\ \bibinfo {author}
  {\bibfnamefont {H.}~\bibnamefont {Schweizer}},\ }\href
  {http://link.aps.org/doi/10.1103/PhysRevLett.86.147} {\bibfield  {journal}
  {\bibinfo  {journal} {Phys. Rev. Lett.},\ }\textbf {\bibinfo {volume} {86}},\
  \bibinfo {pages} {147} (\bibinfo {year} {2001})}\BibitemShut {NoStop}%
\bibitem [{\citenamefont {Dean}\ \emph {et~al.}(2013)\citenamefont {Dean},
  \citenamefont {Wang}, \citenamefont {Maher}, \citenamefont {Forsythe},
  \citenamefont {Ghahari}, \citenamefont {Gao}, \citenamefont {Katoch},
  \citenamefont {Ishigami}, \citenamefont {Moon}, \citenamefont {Koshino},
  \citenamefont {Taniguchi}, \citenamefont {Watanabe}, \citenamefont {Shepard},
  \citenamefont {Hone},\ and\ \citenamefont {Kim}}]{Dean:2013bv}%
  \BibitemOpen
  \bibfield  {author} {\bibinfo {author} {\bibfnamefont {C.~R.}\ \bibnamefont
  {Dean}}, \bibinfo {author} {\bibfnamefont {L.}~\bibnamefont {Wang}}, \bibinfo
  {author} {\bibfnamefont {P.}~\bibnamefont {Maher}}, \bibinfo {author}
  {\bibfnamefont {C.}~\bibnamefont {Forsythe}}, \bibinfo {author}
  {\bibfnamefont {F.}~\bibnamefont {Ghahari}}, \bibinfo {author} {\bibfnamefont
  {Y.}~\bibnamefont {Gao}}, \bibinfo {author} {\bibfnamefont {J.}~\bibnamefont
  {Katoch}}, \bibinfo {author} {\bibfnamefont {M.}~\bibnamefont {Ishigami}},
  \bibinfo {author} {\bibfnamefont {P.}~\bibnamefont {Moon}}, \bibinfo {author}
  {\bibfnamefont {M.}~\bibnamefont {Koshino}}, \bibinfo {author} {\bibfnamefont
  {T.}~\bibnamefont {Taniguchi}}, \bibinfo {author} {\bibfnamefont
  {K.}~\bibnamefont {Watanabe}}, \bibinfo {author} {\bibfnamefont {K.~L.}\
  \bibnamefont {Shepard}}, \bibinfo {author} {\bibfnamefont {J.}~\bibnamefont
  {Hone}}, \ and\ \bibinfo {author} {\bibfnamefont {P.}~\bibnamefont {Kim}},\
  }\href {http://www.nature.com/doifinder/10.1038/nature12186} {\bibfield
  {journal} {\bibinfo  {journal} {Nature},\ }\textbf {\bibinfo {volume}
  {497}},\ \bibinfo {pages} {598} (\bibinfo {year} {2013})}\BibitemShut
  {NoStop}%
\bibitem [{\citenamefont {Hunt}\ \emph {et~al.}(2013)\citenamefont {Hunt},
  \citenamefont {Sanchez-Yamagishi}, \citenamefont {Young}, \citenamefont
  {Yankowitz}, \citenamefont {LeRoy}, \citenamefont {Watanabe}, \citenamefont
  {Taniguchi}, \citenamefont {Moon}, \citenamefont {Koshino}, \citenamefont
  {Jarillo-Herrero},\ and\ \citenamefont {Ashoori}}]{Hunt:2013ef}%
  \BibitemOpen
  \bibfield  {author} {\bibinfo {author} {\bibfnamefont {B.}~\bibnamefont
  {Hunt}}, \bibinfo {author} {\bibfnamefont {J.~D.}\ \bibnamefont
  {Sanchez-Yamagishi}}, \bibinfo {author} {\bibfnamefont {A.~F.}\ \bibnamefont
  {Young}}, \bibinfo {author} {\bibfnamefont {M.}~\bibnamefont {Yankowitz}},
  \bibinfo {author} {\bibfnamefont {B.~J.}\ \bibnamefont {LeRoy}}, \bibinfo
  {author} {\bibfnamefont {K.}~\bibnamefont {Watanabe}}, \bibinfo {author}
  {\bibfnamefont {T.}~\bibnamefont {Taniguchi}}, \bibinfo {author}
  {\bibfnamefont {P.}~\bibnamefont {Moon}}, \bibinfo {author} {\bibfnamefont
  {M.}~\bibnamefont {Koshino}}, \bibinfo {author} {\bibfnamefont
  {P.}~\bibnamefont {Jarillo-Herrero}}, \ and\ \bibinfo {author} {\bibfnamefont
  {R.~C.}\ \bibnamefont {Ashoori}},\ }\href
  {http://www.sciencemag.org/cgi/doi/10.1126/science.1237240} {\bibfield
  {journal} {\bibinfo  {journal} {Science},\ }\textbf {\bibinfo {volume}
  {340}},\ \bibinfo {pages} {1427} (\bibinfo {year} {2013})}\BibitemShut
  {NoStop}%
\bibitem [{\citenamefont {Ponomarenko}\ \emph {et~al.}(2013)\citenamefont
  {Ponomarenko}, \citenamefont {Gorbachev}, \citenamefont {Yu}, \citenamefont
  {Elias}, \citenamefont {Jalil}, \citenamefont {Patel}, \citenamefont
  {Mishchenko}, \citenamefont {Mayorov}, \citenamefont {Woods}, \citenamefont
  {Wallbank}, \citenamefont {Mucha-Kruczynski}, \citenamefont {Piot},
  \citenamefont {Potemski}, \citenamefont {Grigorieva}, \citenamefont
  {Novoselov}, \citenamefont {Guinea}, \citenamefont {Fal'ko},\ and\
  \citenamefont {Geim}}]{Ponomarenko:2013hlb}%
  \BibitemOpen
  \bibfield  {author} {\bibinfo {author} {\bibfnamefont {L.~A.}\ \bibnamefont
  {Ponomarenko}}, \bibinfo {author} {\bibfnamefont {R.~V.}\ \bibnamefont
  {Gorbachev}}, \bibinfo {author} {\bibfnamefont {G.~L.}\ \bibnamefont {Yu}},
  \bibinfo {author} {\bibfnamefont {D.~C.}\ \bibnamefont {Elias}}, \bibinfo
  {author} {\bibfnamefont {R.}~\bibnamefont {Jalil}}, \bibinfo {author}
  {\bibfnamefont {A.~A.}\ \bibnamefont {Patel}}, \bibinfo {author}
  {\bibfnamefont {A.}~\bibnamefont {Mishchenko}}, \bibinfo {author}
  {\bibfnamefont {A.~S.}\ \bibnamefont {Mayorov}}, \bibinfo {author}
  {\bibfnamefont {C.~R.}\ \bibnamefont {Woods}}, \bibinfo {author}
  {\bibfnamefont {J.~R.}\ \bibnamefont {Wallbank}}, \bibinfo {author}
  {\bibfnamefont {M.}~\bibnamefont {Mucha-Kruczynski}}, \bibinfo {author}
  {\bibfnamefont {B.~A.}\ \bibnamefont {Piot}}, \bibinfo {author}
  {\bibfnamefont {M.}~\bibnamefont {Potemski}}, \bibinfo {author}
  {\bibfnamefont {I.~V.}\ \bibnamefont {Grigorieva}}, \bibinfo {author}
  {\bibfnamefont {K.~S.}\ \bibnamefont {Novoselov}}, \bibinfo {author}
  {\bibfnamefont {F.}~\bibnamefont {Guinea}}, \bibinfo {author} {\bibfnamefont
  {V.~I.}\ \bibnamefont {Fal'ko}}, \ and\ \bibinfo {author} {\bibfnamefont
  {A.~K.}\ \bibnamefont {Geim}},\ }\href
  {http://dx.doi.org/10.1038/nature12187} {\bibfield  {journal} {\bibinfo
  {journal} {Nature},\ }\textbf {\bibinfo {volume} {497}},\ \bibinfo {pages}
  {594} (\bibinfo {year} {2013})}\BibitemShut {NoStop}%
\bibitem [{\citenamefont {Thouless}\ \emph {et~al.}(1982)\citenamefont
  {Thouless}, \citenamefont {Kohmoto}, \citenamefont {Nightingale},\ and\
  \citenamefont {den Nijs}}]{Anonymous:4vo2mOrm}%
  \BibitemOpen
  \bibfield  {author} {\bibinfo {author} {\bibfnamefont {D.~J.}\ \bibnamefont
  {Thouless}}, \bibinfo {author} {\bibfnamefont {M.}~\bibnamefont {Kohmoto}},
  \bibinfo {author} {\bibfnamefont {M.~P.}\ \bibnamefont {Nightingale}}, \ and\
  \bibinfo {author} {\bibfnamefont {M.}~\bibnamefont {den Nijs}},\ }\href
  {http://link.aps.org/doi/10.1103/PhysRevLett.49.405} {\bibfield  {journal}
  {\bibinfo  {journal} {Phys. Rev. Lett.},\ }\textbf {\bibinfo {volume} {49}},\
  \bibinfo {pages} {405} (\bibinfo {year} {1982})}\BibitemShut {NoStop}%
\bibitem [{\citenamefont {Osadchy}\ and\ \citenamefont
  {Avron}(2001)}]{Osadchy:2001jm}%
  \BibitemOpen
  \bibfield  {author} {\bibinfo {author} {\bibfnamefont {D.}~\bibnamefont
  {Osadchy}}\ and\ \bibinfo {author} {\bibfnamefont {J.~E.}\ \bibnamefont
  {Avron}},\ }\href {http://link.aip.org/link/JMAPAQ/v42/i12/p5665/s1&Agg=doi}
  {\bibfield  {journal} {\bibinfo  {journal} {J. Math. Phys.},\ }\textbf
  {\bibinfo {volume} {42}},\ \bibinfo {pages} {5665} (\bibinfo {year}
  {2001})}\BibitemShut {NoStop}%
\bibitem [{\citenamefont {Mueller}(2004)}]{Mueller:2004hc}%
  \BibitemOpen
  \bibfield  {author} {\bibinfo {author} {\bibfnamefont {E.}~\bibnamefont
  {Mueller}},\ }\href {http://link.aps.org/doi/10.1103/PhysRevA.70.041603}
  {\bibfield  {journal} {\bibinfo  {journal} {Phys. Rev. A},\ }\textbf
  {\bibinfo {volume} {70}},\ \bibinfo {pages} {041603} (\bibinfo {year}
  {2004})}\BibitemShut {NoStop}%
\bibitem [{\citenamefont {Umucalilar}\ \emph {et~al.}(2008)\citenamefont
  {Umucalilar}, \citenamefont {Zhai},\ and\ \citenamefont
  {Oktel}}]{Umucallar:2008fq}%
  \BibitemOpen
  \bibfield  {author} {\bibinfo {author} {\bibfnamefont {R.~O.}\ \bibnamefont
  {Umucalilar}}, \bibinfo {author} {\bibfnamefont {H.}~\bibnamefont {Zhai}}, \
  and\ \bibinfo {author} {\bibfnamefont {M.}~\bibnamefont {Oktel}},\ }\href
  {http://link.aps.org/doi/10.1103/PhysRevLett.100.070402} {\bibfield
  {journal} {\bibinfo  {journal} {Phys. Rev. Lett.},\ }\textbf {\bibinfo
  {volume} {100}},\ \bibinfo {pages} {070402} (\bibinfo {year}
  {2008})}\BibitemShut {NoStop}%
\bibitem [{\citenamefont {Gerbier}\ and\ \citenamefont
  {Dalibard}(2010)}]{Gerbier:2010ho}%
  \BibitemOpen
  \bibfield  {author} {\bibinfo {author} {\bibfnamefont {F.}~\bibnamefont
  {Gerbier}}\ and\ \bibinfo {author} {\bibfnamefont {J.}~\bibnamefont
  {Dalibard}},\ }\href
  {http://stacks.iop.org/1367-2630/12/i=3/a=033007?key=crossref.84d86d196c336fbca6a62ee5da5b5f59}
  {\bibfield  {journal} {\bibinfo  {journal} {New J. Phys.},\ }\textbf
  {\bibinfo {volume} {12}},\ \bibinfo {pages} {033007} (\bibinfo {year}
  {2010})}\BibitemShut {NoStop}%
\bibitem [{\citenamefont {Kolovsky}(2011)}]{Kolovsky:2011dv}%
  \BibitemOpen
  \bibfield  {author} {\bibinfo {author} {\bibfnamefont {A.~R.}\ \bibnamefont
  {Kolovsky}},\ }\href
  {http://stacks.iop.org/0295-5075/93/i=2/a=20003?key=crossref.42732dd09e62ed27a48c2427229e57be}
  {\bibfield  {journal} {\bibinfo  {journal} {EPL},\ }\textbf {\bibinfo
  {volume} {93}},\ \bibinfo {pages} {20003} (\bibinfo {year}
  {2011})}\BibitemShut {NoStop}%
\bibitem [{\citenamefont {Creffield}\ and\ \citenamefont
  {Sols}(2013)}]{Creffield:2013gp}%
  \BibitemOpen
  \bibfield  {author} {\bibinfo {author} {\bibfnamefont {C.~E.}\ \bibnamefont
  {Creffield}}\ and\ \bibinfo {author} {\bibfnamefont {F.}~\bibnamefont
  {Sols}},\ }\href
  {http://stacks.iop.org/0295-5075/101/i=4/a=40001?key=crossref.30eb8f0daab290b6322c01176f36087d}
  {\bibfield  {journal} {\bibinfo  {journal} {EPL},\ }\textbf {\bibinfo
  {volume} {101}},\ \bibinfo {pages} {40001} (\bibinfo {year}
  {2013})}\BibitemShut {NoStop}%
\bibitem [{\citenamefont {Aidelsburger}\ \emph {et~al.}(2011)\citenamefont
  {Aidelsburger}, \citenamefont {Atala}, \citenamefont {Nascimb{\`e}ne},
  \citenamefont {Trotzky}, \citenamefont {Chen},\ and\ \citenamefont
  {Bloch}}]{Aidelsburger:2011hl}%
  \BibitemOpen
  \bibfield  {author} {\bibinfo {author} {\bibfnamefont {M.}~\bibnamefont
  {Aidelsburger}}, \bibinfo {author} {\bibfnamefont {M.}~\bibnamefont {Atala}},
  \bibinfo {author} {\bibfnamefont {S.}~\bibnamefont {Nascimb{\`e}ne}},
  \bibinfo {author} {\bibfnamefont {S.}~\bibnamefont {Trotzky}}, \bibinfo
  {author} {\bibfnamefont {Y.~A.}\ \bibnamefont {Chen}}, \ and\ \bibinfo
  {author} {\bibfnamefont {I.}~\bibnamefont {Bloch}},\ }\href
  {http://link.aps.org/doi/10.1103/PhysRevLett.107.255301} {\bibfield
  {journal} {\bibinfo  {journal} {Phys. Rev. Lett.},\ }\textbf {\bibinfo
  {volume} {107}},\ \bibinfo {pages} {255301} (\bibinfo {year}
  {2011})}\BibitemShut {NoStop}%
\bibitem [{\citenamefont {Aidelsburger}\ \emph
  {et~al.}(2013){\natexlab{a}}\citenamefont {Aidelsburger}, \citenamefont
  {Atala}, \citenamefont {Nascimb{\`e}ne}, \citenamefont {Trotzky},
  \citenamefont {Chen},\ and\ \citenamefont {Bloch}}]{Aidelsburger:2013du}%
  \BibitemOpen
  \bibfield  {author} {\bibinfo {author} {\bibfnamefont {M.}~\bibnamefont
  {Aidelsburger}}, \bibinfo {author} {\bibfnamefont {M.}~\bibnamefont {Atala}},
  \bibinfo {author} {\bibfnamefont {S.}~\bibnamefont {Nascimb{\`e}ne}},
  \bibinfo {author} {\bibfnamefont {S.}~\bibnamefont {Trotzky}}, \bibinfo
  {author} {\bibfnamefont {Y.~A.}\ \bibnamefont {Chen}}, \ and\ \bibinfo
  {author} {\bibfnamefont {I.}~\bibnamefont {Bloch}},\ }\href
  {http://link.springer.com/10.1007/s00340-013-5418-1} {\bibfield  {journal}
  {\bibinfo  {journal} {Appl. Phys. B}} (\bibinfo {year}
  {2013}{\natexlab{a}})}\BibitemShut {NoStop}%
\bibitem [{\citenamefont {Jaksch}\ and\ \citenamefont
  {Zoller}(2003)}]{Jaksch:2003gd}%
  \BibitemOpen
  \bibfield  {author} {\bibinfo {author} {\bibfnamefont {D.}~\bibnamefont
  {Jaksch}}\ and\ \bibinfo {author} {\bibfnamefont {P.}~\bibnamefont
  {Zoller}},\ }\href {http://iopscience.iop.org/1367-2630/5/1/356} {\bibfield
  {journal} {\bibinfo  {journal} {New J. Phys.},\ }\textbf {\bibinfo {volume}
  {5}},\ \bibinfo {pages} {56} (\bibinfo {year} {2003})}\BibitemShut {NoStop}%
\bibitem [{\citenamefont {Aidelsburger}\ \emph
  {et~al.}(2013){\natexlab{b}}\citenamefont {Aidelsburger}, \citenamefont
  {Atala}, \citenamefont {Lohse}, \citenamefont {Barreiro}, \citenamefont
  {Paredes},\ and\ \citenamefont {Bloch}}]{Aidelsburger:2013ti}%
  \BibitemOpen
  \bibfield  {author} {\bibinfo {author} {\bibfnamefont {M.}~\bibnamefont
  {Aidelsburger}}, \bibinfo {author} {\bibfnamefont {M.}~\bibnamefont {Atala}},
  \bibinfo {author} {\bibfnamefont {M.}~\bibnamefont {Lohse}}, \bibinfo
  {author} {\bibfnamefont {J.~T.}\ \bibnamefont {Barreiro}}, \bibinfo {author}
  {\bibfnamefont {B.}~\bibnamefont {Paredes}}, \ and\ \bibinfo {author}
  {\bibfnamefont {I.}~\bibnamefont {Bloch}},\ }\href
  {http://arxiv.org/abs/1308.0321v1} {\bibfield  {journal} {\bibinfo  {journal}
  {arXiv},\ }\textbf {\bibinfo {volume} {1308.0321}} (\bibinfo {year}
  {2013}{\natexlab{b}})}\BibitemShut {NoStop}%
\bibitem [{\citenamefont {Miyake}\ \emph {et~al.}(2013)\citenamefont {Miyake},
  \citenamefont {Siviloglou}, \citenamefont {Kennedy}, \citenamefont {Burton},\
  and\ \citenamefont {Ketterle}}]{Miyake:2013vbc}%
  \BibitemOpen
  \bibfield  {author} {\bibinfo {author} {\bibfnamefont {H.}~\bibnamefont
  {Miyake}}, \bibinfo {author} {\bibfnamefont {G.~A.}\ \bibnamefont
  {Siviloglou}}, \bibinfo {author} {\bibfnamefont {C.~J.}\ \bibnamefont
  {Kennedy}}, \bibinfo {author} {\bibfnamefont {W.~C.}\ \bibnamefont {Burton}},
  \ and\ \bibinfo {author} {\bibfnamefont {W.}~\bibnamefont {Ketterle}},\
  }\href {http://arxiv.org/abs/1308.1431v3} {\bibfield  {journal} {\bibinfo
  {journal} {arXiv},\ }\textbf {\bibinfo {volume} {1308.1431}} (\bibinfo {year}
  {2013})}\BibitemShut {NoStop}%
\bibitem [{\citenamefont {Alba}\ \emph {et~al.}(2011)\citenamefont {Alba},
  \citenamefont {Fernandez-Gonzalvo}, \citenamefont {Mur-Petit}, \citenamefont
  {Pachos},\ and\ \citenamefont {Garcia-Ripoll}}]{Alba:2011cb}%
  \BibitemOpen
  \bibfield  {author} {\bibinfo {author} {\bibfnamefont {E.}~\bibnamefont
  {Alba}}, \bibinfo {author} {\bibfnamefont {X.}~\bibnamefont
  {Fernandez-Gonzalvo}}, \bibinfo {author} {\bibfnamefont {J.}~\bibnamefont
  {Mur-Petit}}, \bibinfo {author} {\bibfnamefont {J.~K.}\ \bibnamefont
  {Pachos}}, \ and\ \bibinfo {author} {\bibfnamefont {J.~J.}\ \bibnamefont
  {Garcia-Ripoll}},\ }\Doi {10.1103/PhysRevLett.107.235301} {\bibfield
  {journal} {\bibinfo  {journal} {Phys. Rev. Lett.},\ }\textbf {\bibinfo
  {volume} {107}},\ \bibinfo {pages} {235301} (\bibinfo {year}
  {2011})}\BibitemShut {NoStop}%
\bibitem [{\citenamefont {Goldman}\ \emph {et~al.}(2013)\citenamefont
  {Goldman}, \citenamefont {Dalibard}, \citenamefont {Dauphin}, \citenamefont
  {Gerbier}, \citenamefont {Lewenstein}, \citenamefont {Zoller},\ and\
  \citenamefont {Spielman}}]{Goldman:2013dg}%
  \BibitemOpen
  \bibfield  {author} {\bibinfo {author} {\bibfnamefont {N.}~\bibnamefont
  {Goldman}}, \bibinfo {author} {\bibfnamefont {J.}~\bibnamefont {Dalibard}},
  \bibinfo {author} {\bibfnamefont {A.}~\bibnamefont {Dauphin}}, \bibinfo
  {author} {\bibfnamefont {F.}~\bibnamefont {Gerbier}}, \bibinfo {author}
  {\bibfnamefont {M.}~\bibnamefont {Lewenstein}}, \bibinfo {author}
  {\bibfnamefont {P.}~\bibnamefont {Zoller}}, \ and\ \bibinfo {author}
  {\bibfnamefont {I.~B.}\ \bibnamefont {Spielman}},\ }\href
  {http://www.pnas.org/content/110/17/6736.short} {\bibfield  {journal}
  {\bibinfo  {journal} {Proceedings of the National Academy of Sciences of the
  United States of America},\ }\textbf {\bibinfo {volume} {110}},\ \bibinfo
  {pages} {6736} (\bibinfo {year} {2013})}\BibitemShut {NoStop}%
\bibitem [{\citenamefont {Abanin}\ \emph {et~al.}(2013)\citenamefont {Abanin},
  \citenamefont {Kitagawa}, \citenamefont {Bloch},\ and\ \citenamefont
  {Demler}}]{Anonymous:2013cs}%
  \BibitemOpen
  \bibfield  {author} {\bibinfo {author} {\bibfnamefont {D.~A.}\ \bibnamefont
  {Abanin}}, \bibinfo {author} {\bibfnamefont {T.}~\bibnamefont {Kitagawa}},
  \bibinfo {author} {\bibfnamefont {I.}~\bibnamefont {Bloch}}, \ and\ \bibinfo
  {author} {\bibfnamefont {E.}~\bibnamefont {Demler}},\ }\href
  {http://link.aps.org/doi/10.1103/PhysRevLett.110.165304} {\bibfield
  {journal} {\bibinfo  {journal} {Phys. Rev. Lett.},\ }\textbf {\bibinfo
  {volume} {110}},\ \bibinfo {pages} {165304} (\bibinfo {year}
  {2013})}\BibitemShut {NoStop}%
\bibitem [{\citenamefont {Wang}\ \emph {et~al.}(2013)\citenamefont {Wang},
  \citenamefont {Soluyanov},\ and\ \citenamefont {Troyer}}]{Anonymous:2013jga}%
  \BibitemOpen
  \bibfield  {author} {\bibinfo {author} {\bibfnamefont {L.}~\bibnamefont
  {Wang}}, \bibinfo {author} {\bibfnamefont {A.~A.}\ \bibnamefont {Soluyanov}},
  \ and\ \bibinfo {author} {\bibfnamefont {M.}~\bibnamefont {Troyer}},\ }\href
  {http://link.aps.org/doi/10.1103/PhysRevLett.110.166802} {\bibfield
  {journal} {\bibinfo  {journal} {Phys. Rev. Lett.},\ }\textbf {\bibinfo
  {volume} {110}},\ \bibinfo {pages} {166802} (\bibinfo {year}
  {2013})}\BibitemShut {NoStop}%
\bibitem [{\citenamefont {Liu}\ \emph {et~al.}(2013)\citenamefont {Liu},
  \citenamefont {Law}, \citenamefont {Ng},\ and\ \citenamefont
  {Lee}}]{Liu:2013wy}%
  \BibitemOpen
  \bibfield  {author} {\bibinfo {author} {\bibfnamefont {X.-J.}\ \bibnamefont
  {Liu}}, \bibinfo {author} {\bibfnamefont {K.~T.}\ \bibnamefont {Law}},
  \bibinfo {author} {\bibfnamefont {T.~K.}\ \bibnamefont {Ng}}, \ and\ \bibinfo
  {author} {\bibfnamefont {P.~A.}\ \bibnamefont {Lee}},\ }\href
  {http://arxiv.org/abs/1306.5223v2} {\bibfield  {journal} {\bibinfo  {journal}
  {arXiv},\ }\textbf {\bibinfo {volume} {1306.5223}} (\bibinfo {year}
  {2013})}\BibitemShut {NoStop}%
\bibitem [{\citenamefont {Jarrell}\ and\ \citenamefont
  {Gubernatis}(1996)}]{Jarrell:1996uo}%
  \BibitemOpen
  \bibfield  {author} {\bibinfo {author} {\bibfnamefont {M.}~\bibnamefont
  {Jarrell}}\ and\ \bibinfo {author} {\bibfnamefont {J.~E.}\ \bibnamefont
  {Gubernatis}},\ }\href
  {http://www.sciencedirect.com/science/article/pii/0370157395000747}
  {\bibfield  {journal} {\bibinfo  {journal} {Physics Reports},\ }\textbf
  {\bibinfo {volume} {269}},\ \bibinfo {pages} {133} (\bibinfo {year}
  {1996})}\BibitemShut {NoStop}%
\bibitem [{\citenamefont {Hasegawa}\ \emph {et~al.}(1989)\citenamefont
  {Hasegawa}, \citenamefont {Lederer}, \citenamefont {Rice},\ and\
  \citenamefont {Wiegmann}}]{Hasegawa:1989wq}%
  \BibitemOpen
  \bibfield  {author} {\bibinfo {author} {\bibfnamefont {Y.}~\bibnamefont
  {Hasegawa}}, \bibinfo {author} {\bibfnamefont {P.}~\bibnamefont {Lederer}},
  \bibinfo {author} {\bibfnamefont {T.~M.}\ \bibnamefont {Rice}}, \ and\
  \bibinfo {author} {\bibfnamefont {P.~B.}\ \bibnamefont {Wiegmann}},\ }\href
  {http://prl.aps.org/abstract/PRL/v63/i8/p907_1} {\bibfield  {journal}
  {\bibinfo  {journal} {Phys. Rev. Lett.},\ }\textbf {\bibinfo {volume} {63}},\
  \bibinfo {pages} {907} (\bibinfo {year} {1989})}\BibitemShut {NoStop}%
\bibitem [{\citenamefont {Xu}\ \emph {et~al.}(2008)\citenamefont {Xu},
  \citenamefont {Yang}, \citenamefont {Qin},\ and\ \citenamefont
  {Xiang}}]{Xu:2008gq}%
  \BibitemOpen
  \bibfield  {author} {\bibinfo {author} {\bibfnamefont {W.}~\bibnamefont
  {Xu}}, \bibinfo {author} {\bibfnamefont {L.}~\bibnamefont {Yang}}, \bibinfo
  {author} {\bibfnamefont {M.}~\bibnamefont {Qin}}, \ and\ \bibinfo {author}
  {\bibfnamefont {T.}~\bibnamefont {Xiang}},\ }\href
  {http://link.aps.org/doi/10.1103/PhysRevB.78.241102} {\bibfield  {journal}
  {\bibinfo  {journal} {Phys. Rev. B},\ }\textbf {\bibinfo {volume} {78}},\
  \bibinfo {pages} {241102} (\bibinfo {year} {2008})}\BibitemShut {NoStop}%
\bibitem [{\citenamefont {Yang}\ \emph {et~al.}(2012)\citenamefont {Yang},
  \citenamefont {Xu}, \citenamefont {Qin},\ and\ \citenamefont
  {Xiang}}]{Yang:2012gf}%
  \BibitemOpen
  \bibfield  {author} {\bibinfo {author} {\bibfnamefont {L.-P.}\ \bibnamefont
  {Yang}}, \bibinfo {author} {\bibfnamefont {W.-H.}\ \bibnamefont {Xu}},
  \bibinfo {author} {\bibfnamefont {M.-P.}\ \bibnamefont {Qin}}, \ and\
  \bibinfo {author} {\bibfnamefont {T.}~\bibnamefont {Xiang}},\ }\href
  {http://jpsj.ipap.jp/link?JPSJ/81/044605/} {\bibfield  {journal} {\bibinfo
  {journal} {J. Phys. Soc. Jpn.},\ }\textbf {\bibinfo {volume} {81}},\ \bibinfo
  {pages} {044605} (\bibinfo {year} {2012})}\BibitemShut {NoStop}%
\bibitem [{\citenamefont {Gemelke}\ \emph {et~al.}(2009)\citenamefont
  {Gemelke}, \citenamefont {Zhang}, \citenamefont {Hung},\ and\ \citenamefont
  {Chin}}]{Gemelke:2009ja}%
  \BibitemOpen
  \bibfield  {author} {\bibinfo {author} {\bibfnamefont {N.}~\bibnamefont
  {Gemelke}}, \bibinfo {author} {\bibfnamefont {X.}~\bibnamefont {Zhang}},
  \bibinfo {author} {\bibfnamefont {C.-L.}\ \bibnamefont {Hung}}, \ and\
  \bibinfo {author} {\bibfnamefont {C.}~\bibnamefont {Chin}},\ }\href
  {http://dx.doi.org/10.1038/nature08244} {\bibfield  {journal} {\bibinfo
  {journal} {Nature},\ }\textbf {\bibinfo {volume} {460}},\ \bibinfo {pages}
  {995} (\bibinfo {year} {2009})}\BibitemShut {NoStop}%
\bibitem [{\citenamefont {Van~Houcke}\ \emph {et~al.}(2012)\citenamefont
  {Van~Houcke}, \citenamefont {Werner}, \citenamefont {Kozik}, \citenamefont
  {Prokof'ev}, \citenamefont {Svistunov}, \citenamefont {Ku}, \citenamefont
  {Sommer}, \citenamefont {Cheuk}, \citenamefont {Schirotzek},\ and\
  \citenamefont {Zwierlein}}]{VanHoucke:2012ic}%
  \BibitemOpen
  \bibfield  {author} {\bibinfo {author} {\bibfnamefont {K.}~\bibnamefont
  {Van~Houcke}}, \bibinfo {author} {\bibfnamefont {F.}~\bibnamefont {Werner}},
  \bibinfo {author} {\bibfnamefont {E.}~\bibnamefont {Kozik}}, \bibinfo
  {author} {\bibfnamefont {N.}~\bibnamefont {Prokof'ev}}, \bibinfo {author}
  {\bibfnamefont {B.}~\bibnamefont {Svistunov}}, \bibinfo {author}
  {\bibfnamefont {M.~J.~H.}\ \bibnamefont {Ku}}, \bibinfo {author}
  {\bibfnamefont {A.~T.}\ \bibnamefont {Sommer}}, \bibinfo {author}
  {\bibfnamefont {L.~W.}\ \bibnamefont {Cheuk}}, \bibinfo {author}
  {\bibfnamefont {A.}~\bibnamefont {Schirotzek}}, \ and\ \bibinfo {author}
  {\bibfnamefont {M.~W.}\ \bibnamefont {Zwierlein}},\ }\href
  {http://www.nature.com/doifinder/10.1038/nphys2273} {\bibfield  {journal}
  {\bibinfo  {journal} {Nature Physics},\ }\textbf {\bibinfo {volume} {8}},\
  \bibinfo {pages} {366} (\bibinfo {year} {2012})}\BibitemShut {NoStop}%
\bibitem [{\citenamefont {Ku}\ \emph {et~al.}(2012)\citenamefont {Ku},
  \citenamefont {Sommer}, \citenamefont {Cheuk},\ and\ \citenamefont
  {Zwierlein}}]{Ku:2012gra}%
  \BibitemOpen
  \bibfield  {author} {\bibinfo {author} {\bibfnamefont {M.~J.~H.}\
  \bibnamefont {Ku}}, \bibinfo {author} {\bibfnamefont {A.~T.}\ \bibnamefont
  {Sommer}}, \bibinfo {author} {\bibfnamefont {L.~W.}\ \bibnamefont {Cheuk}}, \
  and\ \bibinfo {author} {\bibfnamefont {M.~W.}\ \bibnamefont {Zwierlein}},\
  }\href {http://www.sciencemag.org/cgi/doi/10.1126/science.1214987} {\bibfield
   {journal} {\bibinfo  {journal} {Science},\ }\textbf {\bibinfo {volume}
  {335}},\ \bibinfo {pages} {563} (\bibinfo {year} {2012})}\BibitemShut
  {NoStop}%
\bibitem [{\citenamefont {Lee}\ \emph {et~al.}(2012)\citenamefont {Lee},
  \citenamefont {Heo}, \citenamefont {Choi}, \citenamefont {Wang},
  \citenamefont {Christensen}, \citenamefont {Rvachov},\ and\ \citenamefont
  {Ketterle}}]{Lee:2012jm}%
  \BibitemOpen
  \bibfield  {author} {\bibinfo {author} {\bibfnamefont {Y.-R.}\ \bibnamefont
  {Lee}}, \bibinfo {author} {\bibfnamefont {M.-S.}\ \bibnamefont {Heo}},
  \bibinfo {author} {\bibfnamefont {J.-H.}\ \bibnamefont {Choi}}, \bibinfo
  {author} {\bibfnamefont {T.~T.}\ \bibnamefont {Wang}}, \bibinfo {author}
  {\bibfnamefont {C.~A.}\ \bibnamefont {Christensen}}, \bibinfo {author}
  {\bibfnamefont {T.~M.}\ \bibnamefont {Rvachov}}, \ and\ \bibinfo {author}
  {\bibfnamefont {W.}~\bibnamefont {Ketterle}},\ }\href
  {http://link.aps.org/doi/10.1103/PhysRevA.85.063615} {\bibfield  {journal}
  {\bibinfo  {journal} {Phys. Rev. A},\ }\textbf {\bibinfo {volume} {85}},\
  \bibinfo {pages} {063615} (\bibinfo {year} {2012})}\BibitemShut {NoStop}%
\bibitem [{\citenamefont {McKay}\ and\ \citenamefont
  {DeMarco}(2011)}]{McKay:2011fc}%
  \BibitemOpen
  \bibfield  {author} {\bibinfo {author} {\bibfnamefont {D.~C.}\ \bibnamefont
  {McKay}}\ and\ \bibinfo {author} {\bibfnamefont {B.}~\bibnamefont
  {DeMarco}},\ }\href
  {http://stacks.iop.org/0034-4885/74/i=5/a=054401?key=crossref.b6c5a8abc6cac56fe2b79243d6632ad5}
  {\bibfield  {journal} {\bibinfo  {journal} {Rep. Prog. Phys.},\ }\textbf
  {\bibinfo {volume} {74}},\ \bibinfo {pages} {054401} (\bibinfo {year}
  {2011})}\BibitemShut {NoStop}%
\bibitem [{\citenamefont {Fuchs}\ \emph {et~al.}(2011)\citenamefont {Fuchs},
  \citenamefont {Gull}, \citenamefont {Pollet}, \citenamefont {Burovski},
  \citenamefont {Kozik}, \citenamefont {Pruschke},\ and\ \citenamefont
  {Troyer}}]{Fuchs:2011cha}%
  \BibitemOpen
  \bibfield  {author} {\bibinfo {author} {\bibfnamefont {S.}~\bibnamefont
  {Fuchs}}, \bibinfo {author} {\bibfnamefont {E.}~\bibnamefont {Gull}},
  \bibinfo {author} {\bibfnamefont {L.}~\bibnamefont {Pollet}}, \bibinfo
  {author} {\bibfnamefont {E.}~\bibnamefont {Burovski}}, \bibinfo {author}
  {\bibfnamefont {E.}~\bibnamefont {Kozik}}, \bibinfo {author} {\bibfnamefont
  {T.}~\bibnamefont {Pruschke}}, \ and\ \bibinfo {author} {\bibfnamefont
  {M.}~\bibnamefont {Troyer}},\ }\href
  {http://link.aps.org/doi/10.1103/PhysRevLett.106.030401} {\bibfield
  {journal} {\bibinfo  {journal} {Phys. Rev. Lett.},\ }\textbf {\bibinfo
  {volume} {106}},\ \bibinfo {pages} {030401} (\bibinfo {year}
  {2011})}\BibitemShut {NoStop}%
\bibitem [{\citenamefont {Sandvik}(1998)}]{Sandvik:1998ut}%
  \BibitemOpen
  \bibfield  {author} {\bibinfo {author} {\bibfnamefont {A.~W.}\ \bibnamefont
  {Sandvik}},\ }\href {http://prb.aps.org/abstract/PRB/v57/i17/p10287_1}
  {\bibfield  {journal} {\bibinfo  {journal} {Phys. Rev. B},\ }\textbf
  {\bibinfo {volume} {57}},\ \bibinfo {pages} {10287} (\bibinfo {year}
  {1998})}\BibitemShut {NoStop}%
\bibitem [{\citenamefont {Mishchenko}\ \emph {et~al.}(2000)\citenamefont
  {Mishchenko}, \citenamefont {Prokof'ev}, \citenamefont {Sakamoto},\ and\
  \citenamefont {{Svistunov, BV}}}]{Mishchenko:2000vm}%
  \BibitemOpen
  \bibfield  {author} {\bibinfo {author} {\bibfnamefont {A.~S.}\ \bibnamefont
  {Mishchenko}}, \bibinfo {author} {\bibfnamefont {N.~V.}\ \bibnamefont
  {Prokof'ev}}, \bibinfo {author} {\bibfnamefont {A.}~\bibnamefont {Sakamoto}},
  \ and\ \bibinfo {author} {\bibnamefont {{Svistunov, BV}}},\ }\href
  {http://prb.aps.org/abstract/PRB/v62/i10/p6317_1} {\bibfield  {journal}
  {\bibinfo  {journal} {Phys. Rev. B},\ }\textbf {\bibinfo {volume} {62}},\
  \bibinfo {pages} {6317} (\bibinfo {year} {2000})}\BibitemShut {NoStop}%
\bibitem [{\citenamefont {Beach}(2004)}]{Beach:2004uc}%
  \BibitemOpen
  \bibfield  {author} {\bibinfo {author} {\bibfnamefont {K.~S.~D.}\
  \bibnamefont {Beach}},\ }\href {http://arxiv.org/abs/cond-mat/0403055v1}
  {\bibfield  {journal} {\bibinfo  {journal} {arXiv},\ }\textbf {\bibinfo
  {volume} {cond-mat/0403055}} (\bibinfo {year} {2004})}\BibitemShut {NoStop}%
\bibitem [{\citenamefont {Fuchs}\ \emph {et~al.}(2010)\citenamefont {Fuchs},
  \citenamefont {Pruschke},\ and\ \citenamefont {Jarrell}}]{Fuchs:2010it}%
  \BibitemOpen
  \bibfield  {author} {\bibinfo {author} {\bibfnamefont {S.}~\bibnamefont
  {Fuchs}}, \bibinfo {author} {\bibfnamefont {T.}~\bibnamefont {Pruschke}}, \
  and\ \bibinfo {author} {\bibfnamefont {M.}~\bibnamefont {Jarrell}},\ }\href
  {http://link.aps.org/doi/10.1103/PhysRevE.81.056701} {\bibfield  {journal}
  {\bibinfo  {journal} {Phys. Rev. E},\ }\textbf {\bibinfo {volume} {81}},\
  \bibinfo {pages} {056701} (\bibinfo {year} {2010})}\BibitemShut {NoStop}%
\bibitem [{\citenamefont {Prokof'ev}\ and\ \citenamefont
  {Svistunov}(2013)}]{Boris}%
  \BibitemOpen
  \bibfield  {author} {\bibinfo {author} {\bibfnamefont {N.}~\bibnamefont
  {Prokof'ev}}\ and\ \bibinfo {author} {\bibfnamefont {B.}~\bibnamefont
  {Svistunov}},\ }\href {http://arxiv.org/abs/1304.5198} {\bibfield  {journal}
  {\bibinfo  {journal} {arXiv},\ }\textbf {\bibinfo {volume} {1304.5198}}
  (\bibinfo {year} {2013})}\BibitemShut {NoStop}%
\bibitem [{\citenamefont {Silver}\ \emph {et~al.}(1990)\citenamefont {Silver},
  \citenamefont {Sivia},\ and\ \citenamefont {Gubernatis}}]{Silver1996}%
  \BibitemOpen
  \bibfield  {author} {\bibinfo {author} {\bibfnamefont {R.~N.}\ \bibnamefont
  {Silver}}, \bibinfo {author} {\bibfnamefont {D.~S.}\ \bibnamefont {Sivia}}, \
  and\ \bibinfo {author} {\bibfnamefont {J.~E.}\ \bibnamefont {Gubernatis}},\
  }\href
  {http://adsabs.harvard.edu/cgi-bin/nph-data_query?bibcode=1990PhRvB..41.2380S&link_type=ABSTRACT}
  {\bibfield  {journal} {\bibinfo  {journal} {Phys. Rev. B},\ }\textbf
  {\bibinfo {volume} {41}},\ \bibinfo {pages} {2380} (\bibinfo {year}
  {1990})}\BibitemShut {NoStop}%
\bibitem [{\citenamefont {Zhou}\ and\ \citenamefont {Ho}(2011)}]{Zhou:2011dj}%
  \BibitemOpen
  \bibfield  {author} {\bibinfo {author} {\bibfnamefont {Q.}~\bibnamefont
  {Zhou}}\ and\ \bibinfo {author} {\bibfnamefont {T.-L.}\ \bibnamefont {Ho}},\
  }\href {http://link.aps.org/doi/10.1103/PhysRevLett.106.225301} {\bibfield
  {journal} {\bibinfo  {journal} {Phys. Rev. Lett.},\ }\textbf {\bibinfo
  {volume} {106}},\ \bibinfo {pages} {225301} (\bibinfo {year}
  {2011})}\BibitemShut {NoStop}%
\bibitem [{\citenamefont {Sanner}\ \emph {et~al.}(2010)\citenamefont {Sanner},
  \citenamefont {Su}, \citenamefont {Keshet}, \citenamefont {Gommers},
  \citenamefont {Shin}, \citenamefont {Huang},\ and\ \citenamefont
  {Ketterle}}]{Sanner:2010hq}%
  \BibitemOpen
  \bibfield  {author} {\bibinfo {author} {\bibfnamefont {C.}~\bibnamefont
  {Sanner}}, \bibinfo {author} {\bibfnamefont {E.~J.}\ \bibnamefont {Su}},
  \bibinfo {author} {\bibfnamefont {A.}~\bibnamefont {Keshet}}, \bibinfo
  {author} {\bibfnamefont {R.}~\bibnamefont {Gommers}}, \bibinfo {author}
  {\bibfnamefont {Y.-i.}\ \bibnamefont {Shin}}, \bibinfo {author}
  {\bibfnamefont {W.}~\bibnamefont {Huang}}, \ and\ \bibinfo {author}
  {\bibfnamefont {W.}~\bibnamefont {Ketterle}},\ }\href
  {http://link.aps.org/doi/10.1103/PhysRevLett.105.040402} {\bibfield
  {journal} {\bibinfo  {journal} {Phys. Rev. Lett.},\ }\textbf {\bibinfo
  {volume} {105}},\ \bibinfo {pages} {040402} (\bibinfo {year}
  {2010})}\BibitemShut {NoStop}%
\bibitem [{\citenamefont {M{\"u}ller}\ \emph {et~al.}(2010)\citenamefont
  {M{\"u}ller}, \citenamefont {Zimmermann}, \citenamefont {Meineke},
  \citenamefont {Brantut}, \citenamefont {Esslinger},\ and\ \citenamefont
  {Moritz}}]{Muller:2010ei}%
  \BibitemOpen
  \bibfield  {author} {\bibinfo {author} {\bibfnamefont {T.}~\bibnamefont
  {M{\"u}ller}}, \bibinfo {author} {\bibfnamefont {B.}~\bibnamefont
  {Zimmermann}}, \bibinfo {author} {\bibfnamefont {J.}~\bibnamefont {Meineke}},
  \bibinfo {author} {\bibfnamefont {J.-P.}\ \bibnamefont {Brantut}}, \bibinfo
  {author} {\bibfnamefont {T.}~\bibnamefont {Esslinger}}, \ and\ \bibinfo
  {author} {\bibfnamefont {H.}~\bibnamefont {Moritz}},\ }\href
  {http://link.aps.org/doi/10.1103/PhysRevLett.105.040401} {\bibfield
  {journal} {\bibinfo  {journal} {Phys. Rev. Lett.},\ }\textbf {\bibinfo
  {volume} {105}},\ \bibinfo {pages} {040401} (\bibinfo {year}
  {2010})}\BibitemShut {NoStop}%
\bibitem [{\citenamefont {Ho}\ and\ \citenamefont {Zhou}(2009)}]{Ho:2009cp}%
  \BibitemOpen
  \bibfield  {author} {\bibinfo {author} {\bibfnamefont {T.-L.}\ \bibnamefont
  {Ho}}\ and\ \bibinfo {author} {\bibfnamefont {Q.}~\bibnamefont {Zhou}},\
  }\href {http://www.nature.com/doifinder/10.1038/nphys1477} {\bibfield
  {journal} {\bibinfo  {journal} {Nature Physics},\ }\textbf {\bibinfo {volume}
  {6}},\ \bibinfo {pages} {131} (\bibinfo {year} {2009})}\BibitemShut {NoStop}%
\bibitem [{\citenamefont {Ma}\ \emph {et~al.}(2010)\citenamefont {Ma},
  \citenamefont {Pollet},\ and\ \citenamefont {Troyer}}]{Ma:2010do}%
  \BibitemOpen
  \bibfield  {author} {\bibinfo {author} {\bibfnamefont {P.~N.}\ \bibnamefont
  {Ma}}, \bibinfo {author} {\bibfnamefont {L.}~\bibnamefont {Pollet}}, \ and\
  \bibinfo {author} {\bibfnamefont {M.}~\bibnamefont {Troyer}},\ }\href
  {http://link.aps.org/doi/10.1103/PhysRevA.82.033627} {\bibfield  {journal}
  {\bibinfo  {journal} {Phys. Rev. A},\ }\textbf {\bibinfo {volume} {82}},\
  \bibinfo {pages} {033627} (\bibinfo {year} {2010})}\BibitemShut {NoStop}%
\bibitem [{\citenamefont {Tarruell}\ \emph {et~al.}(2012)\citenamefont
  {Tarruell}, \citenamefont {Greif}, \citenamefont {Uehlinger}, \citenamefont
  {Jotzu},\ and\ \citenamefont {Esslinger}}]{Tarruell:2012db}%
  \BibitemOpen
  \bibfield  {author} {\bibinfo {author} {\bibfnamefont {L.}~\bibnamefont
  {Tarruell}}, \bibinfo {author} {\bibfnamefont {D.}~\bibnamefont {Greif}},
  \bibinfo {author} {\bibfnamefont {T.}~\bibnamefont {Uehlinger}}, \bibinfo
  {author} {\bibfnamefont {G.}~\bibnamefont {Jotzu}}, \ and\ \bibinfo {author}
  {\bibfnamefont {T.}~\bibnamefont {Esslinger}},\ }\href
  {http://www.nature.com/doifinder/10.1038/nature10871} {\bibfield  {journal}
  {\bibinfo  {journal} {Nature},\ }\textbf {\bibinfo {volume} {483}},\ \bibinfo
  {pages} {302} (\bibinfo {year} {2012})}\BibitemShut {NoStop}%
\bibitem [{\citenamefont {Jo}\ \emph {et~al.}(2012)\citenamefont {Jo},
  \citenamefont {Guzman}, \citenamefont {Thomas}, \citenamefont {Hosur},
  \citenamefont {Vishwanath},\ and\ \citenamefont {Stamper-Kurn}}]{Jo:2012br}%
  \BibitemOpen
  \bibfield  {author} {\bibinfo {author} {\bibfnamefont {G.-B.}\ \bibnamefont
  {Jo}}, \bibinfo {author} {\bibfnamefont {J.}~\bibnamefont {Guzman}}, \bibinfo
  {author} {\bibfnamefont {C.~K.}\ \bibnamefont {Thomas}}, \bibinfo {author}
  {\bibfnamefont {P.}~\bibnamefont {Hosur}}, \bibinfo {author} {\bibfnamefont
  {A.}~\bibnamefont {Vishwanath}}, \ and\ \bibinfo {author} {\bibfnamefont
  {D.~M.}\ \bibnamefont {Stamper-Kurn}},\ }\href
  {http://link.aps.org/doi/10.1103/PhysRevLett.108.045305} {\bibfield
  {journal} {\bibinfo  {journal} {Phys. Rev. Lett.},\ }\textbf {\bibinfo
  {volume} {108}},\ \bibinfo {pages} {045305} (\bibinfo {year}
  {2012})}\BibitemShut {NoStop}%
\bibitem [{\citenamefont {Bauer~et al}(2011)}]{BBauer:2011tz}%
  \BibitemOpen
  \bibfield  {author} {\bibinfo {author} {\bibfnamefont {B.}~\bibnamefont
  {Bauer~et al}},\ }\href {http://iopscience.iop.org/1742-5468/2011/05/P05001}
  {\bibfield  {journal} {\bibinfo  {journal} {J. Stat. Mech.: Theor. Exp.},\
  }\textbf {\bibinfo {volume} {2011}},\ \bibinfo {pages} {P05001} (\bibinfo
  {year} {2011})}\BibitemShut {NoStop}%
\end{thebibliography}%
%\putbib[ThermalHofstadter]
%\end{bibunit}

\end{document}